\documentclass[pra,twocolumn,nopacs,superscriptaddress,amssymb]{revtex4}

\usepackage[colorlinks=true,linkcolor=magenta,citecolor=magenta,urlcolor=magenta,linktocpage=true]{hyperref}
\usepackage[utf8]{inputenc}
\usepackage[english]{babel}
\usepackage[T1]{fontenc}
\usepackage{amsmath}
\usepackage{cleveref}
\usepackage{svg}
\usepackage{tabularx}
\usepackage{array, makecell}

\usepackage{listings}
\usepackage{xcolor}
\definecolor{codegreen}{rgb}{0,0.6,0}
\definecolor{codegray}{rgb}{0.5,0.5,0.5}
\definecolor{codepurple}{rgb}{0.58,0,0.82}
\definecolor{backcolour}{rgb}{0.95,0.95,0.92}
\lstdefinestyle{mystyle}{
    backgroundcolor=\color{backcolour},   
    commentstyle=\color{codegreen},
    keywordstyle=\color{magenta},
    numberstyle=\tiny\color{codegray},
    stringstyle=\color{codepurple},
    basicstyle=\ttfamily\footnotesize,
    breakatwhitespace=false,         
    breaklines=true,                 
    captionpos=b,                    
    keepspaces=true,                 
    numbers=none,                    
    numbersep=5pt,                  
    showspaces=false,                
    showstringspaces=false,
    showtabs=false,                  
    tabsize=2
}
\usepackage{psfrag,graphicx}
\usepackage{dcolumn}
\usepackage{bm}
\usepackage{amsfonts,amssymb,amsmath}        
\usepackage{slashed}
\usepackage[utf8]{inputenc}
\usepackage{graphicx}
\usepackage[caption=false]{subfig}
\usepackage{cancel}
\usepackage{xcolor}
\usepackage{amsmath}
\setcitestyle{numbers,square}

\newcommand{\be}{\begin{equation}}
\newcommand{\ee}{\end{equation}}
\newcommand{\bq}{\begin{eqnarray}}
\newcommand{\eq}{\end{eqnarray}}

\makeatletter

\renewenvironment{widetext@grid}{%
  \par\ignorespaces
  \setbox\widetext@top\vbox{%
   \vskip15\p@
   \hb@xt@\hsize{%
    \leaders\hrule\hfil
    \vrule\@height6\p@
   }%
   \vskip6\p@
  }%
  \setbox\widetext@bot\hb@xt@\hsize{%
    \vrule\@depth6\p@
    \leaders\hrule\hfil
  }%
  \onecolumngrid

  \let\set@footnotewidth\set@footnotewidth@ii
}{%
  \par

  \twocolumngrid\global\@ignoretrue
  \@endpetrue
}%

\makeatother

\begin{document}
\title{Principle of least action for quasi-adiabatic state transfers with dissipation}
\author{Si Luo}
\email{north.ls@outlook.com}
\affiliation{Graduate School of China Academy of Engineering Physics, Beijing, 100193, China}
\affiliation{Beijing Computational Science Research Center, Beijing 100094, China}
\author{Yinan Fang}
\affiliation{Beijing Computational Science Research Center, Beijing 100094, China}
\affiliation{School of Physics and Astronomy, Yunnan University, Kunming 650091, China}
\author{Yingdan Wang}
\affiliation{Institute of Theoretical Physics, Chinese Academy of Sciences, Beijing 100190, China}
\author{Stefano Chesi}
\email{stefano.chesi@csrc.ac.cn}
\affiliation{Beijing Computational Science Research Center, Beijing 100094, China}

\date{\today}

\begin{abstract}
We discuss a general formalism to optimize quasi-adiabatic state-transfer protocols, where high-fidelity is achieved by maintaining  the system in a `dark' subspace protected from the dominant dissipative channels. We cast the residual fidelity loss, induced by a combination of dissipation and non-adiabatic transitions, in the form of a classical action where the time-dependent control parameters act as coordinates. This allows us to apply the least action principle, yielding the fidelity upper-bound and the corresponding optimal transfer time. As an application, we analyze a system of two qubits subject to weak relaxation and dephasing, interacting through a strongly dissipative quantum bus. In this case, using our formalism, we obtain a full characterization of the optimal state-transfer fidelity.

\end{abstract}
\maketitle
\vspace{5mm}
\newpage

In the field of quantum information processing, it is often crucial to accurately transfer a quantum state from a ``source''  to a ``target'' system. Among numerous techniques designed to achieve this goal ~\cite{peirce1988,ohtsuki1999,Werschnik2007QuantumOC}, an important class is represented by protocols where the state transfer is achieved by slowly evolving an instantaneous eigenstate. The basic idea of this approach is the same behind stimulated Raman adiabatic passage (STIRAP) ~\cite{gaubatz_population_1988,gaubatz1990population,2001_BOOK_Vitanov,adiabatic_protocols,Shore2008CoherentMO,shore2013,bergmann2015perspective,vitanov2017}, where population transfer in a three-level $\Lambda$-system is realized by applying two coherent pulses in a counter-intuitive sequence. In this manner, the system remains in an instantaneous eigenstate which is dark under the double-resonance condition and avoids the highly lossy intermediate state. Provided that the control acts on a time-scale much longer than the inverse Rabi frequency~\cite{bergmann2015perspective,gaubatz1990population}, the dynamics is adiabatic and robust against small perturbations to the pulses' profiles. 

This type of approach can be generalized to more complex quasi-adiabatic protocols~\cite{greentree2004coherent,ivanov2005spontaneous,goto2008upper,scala2010stimulated,yuan2012controllability,vogt2012influence,hou2013quantum,dupont2015microwave,wang2017optimization,stefanatos2021} which, as in STIRAP, are particularly effective when there is a clear separation $\gamma \gg \kappa$ between the typical decoherence rates associated with `bright' and `dark' subspaces. The application of adiabatic protocols has recently seen successful application to  solid state devices, besides atomic and molecular systems, such as the robust spin manipulation of nitrogen-vacancy (NV) center in diamond~\cite{golter2013nuclear,golter2014optically,golter2014protecting}, and the demonstration of an improved population transfer fidelity in superconducting qubits~\cite{di2016coherent,kumar2016stimulated,xu2016coherent,premaratne2017microwave}.  In these solid-state devices, however, the typical dissipation rate $\kappa$ associated with source and target states may not be negligible on the time-scale of the whole operation process. Therefore, a well designed pulse sequence should carefully balance two competing requirements: Following adiabatically a dark state
and minimizing the transfer time~\cite{wang2017optimization}.

The choice of the transfer transfer time is especially crucial, as  can be understood from the two extreme scenarios. When $\kappa \to 0 $, an ideal state transfer can be realized by the adiabatic evolution, i.e., the optimal transfer time becomes infinitely long. Instead, when approaching the limit when all decoherence rates become comparable, $\gamma \sim \kappa$, it is natural to try to optimize the protocol by making the transfer time as short as possible. To this end, approaches such as shortcut to adiabaticity~\cite{shore1997,bergmann1999,Emmanouilidou2000,demirplak_consistency_2008,Berry_2009,delcampo2013,Torrontegui2013,verdeny2014,baksic2016speeding,ribeiro2017systematic,guery2019,petiziol_optimized_2020,Evangelakos2023} can be of considerable value. Furthermore, the detailed time dependence of the control parameters plays a significant role. Previous studies have explored these issues extensively~\cite{greentree2004coherent,ivanov2005spontaneous,goto2008upper,scala2010stimulated,yuan2012controllability,vogt2012influence,hou2013quantum,dupont2015microwave,wang2017optimization,stefanatos2021} but some limitations remain. For example, constrains on the form of dissipation~\cite{yuan2012controllability}, specific (e.g., Gaussian) ~\cite{greentree2004coherent,ivanov2005spontaneous,goto2008upper,scala2010stimulated,vogt2012influence} or heuristically inspired~\cite{baksic2016speeding} pulse shapes, or studies which rely on direct numerical simulations~\cite{hou2013quantum,dupont2015microwave}.

Here, by considering a generic type of environment~\cite{Petruccione,Rivas2011OpenQS}, we discuss a physically appealing formulation of quasi-adiabatic state transfer protocols which is applicable to the desirable limit (for quantum information processing) of a high fidelity.  Specifically, we cast the leading-order fidelity loss in the form of a classical action~\cite{CMechanics} over the time-dependent control parameters. This provides a mechanical analog to the dissipaive evolution, thus allowing us to optimize the fidelity using the least-action principle. We apply our method to a state transfer between two weakly dissipative qubits interacting with a common (and strongly damped) quantum bus. For this model, we derive a transparent picture of the optimization problem in terms of a 1D particle moving under an external potential, whose specific form is determined by a combination of the various decoherence mechanisms. This formulation also allows us to derive several analytical expressions for the optimal transfer-time and the fidelity upper-bound.

The paper is structured as follows: We start in Sec.~\ref{Variational approach} with a formal discussion of state transfer optimization for an open quantum systems in the adiabatic limit. By writing the leading-order fidelity loss as an action over tunable system  parameters, we express the fidelity optimization problem in terms of a least-action principle.  The formalism is then applied in Sec.~\ref{Model and perturbation} to the state transfer between two dissipative qubits, mediated by a quantum bus. Finally, our conclusions are presented in Sec.~\ref{conclusion} and technical details can be found in the Appendices.

\section{Quasi-Adiabatic Fidelity}\label{Variational approach}
 
The state transfer protocols we consider are realized through the generic Hamiltonian:
\begin{align} \label{eq:genericH}
\hat{H}(t)&=\sum_{j=1}^{N}G_{j}(t)\hat{H}_{j}\;,    
 \end{align}
 where $\hat{H}_{j}$ are time-independent operators. Ideally, the input state $|\psi_{\rm i}\rangle$ evolves to the target state $|\psi_{\rm f}\rangle$ after a transfer time $t_{\rm f}$, but this process is generally impossible to realize exactly, due  to physical constraints (e.g., restrictions on the maximum couplings) and the influence of noise. One is then faced with the problem of finding the optimal time-dependence for the vector ${\bf G}(t)$ of control parameters. Here we analyze this question for the class of state-transfer protocols based on a quasi-adiabatic process, i.e., the ideal state transfer is given by the adiabatic evolution ($\hbar =1$):
 \begin{equation}\label{adiabatic_U}
    \hat{U}_0(t)=\sum_n e^{i\gamma_n(t)-i\phi_n(t)}|E_n(t)\rangle\langle E_n(0)|\;,
\end{equation}
where $|E_n(t)\rangle$ are the instantaneous eigenstates of $\hat{H}(t)$. For simplicity, we will take the $|E_n(t)\rangle$ as non-degenerate. In Eq.~(\ref{adiabatic_U}),  $\phi_n(t)=\int_0^t E_n(\tau)d\tau$ is the dynamical phase and $\gamma_n(t)=i\int_0^t\langle E_n(\tau)|\dot{E}_n(\tau)\rangle d\tau$ is the geometric phase. Without loss of generality we will focus on the eigenstate with $n=1$. Thus, we identify  $|\psi_{\rm i}\rangle = |E_1(0)\rangle $ and  $|\psi_{\rm f}\rangle = |E_1(t_{\rm f})\rangle $.

\subsection{Unitary dynamics}

The first source of imperfection in the state-transfer protocol are non-adiabatic transitions induced by the unitary time evolution, which we characterize with the help of non-adiabatic perturbation theory \cite{Garrison1986, Rigolin2008, DeGrandi2010}. Writing the Hamiltonian as the sum of two terms, $\hat{H}(t)=\hat{H}_0(t)+\hat{H}_{1}(t)$, we require $\hat{H}_0(t)$ to induce the exact adiabatic evolution of Eq.~(\ref{adiabatic_U}), i.e., $\hat{U}_0(t)=\mathcal{T}\exp[-i\int_0^t \hat{H}_0(\tau)d\tau]$ (where $\mathcal{T}$ is the time ordering operator). This requirement determines the non-adiabatic correction as follows:
\begin{equation}\label{H0def}
\hat{H}_1(t)=i\sum_n\sum_{m\neq n}\epsilon_{m,n}(t)\Delta_{m,n}(t)|E_m(t)\rangle\langle E_n(t)|\;, 
\end{equation}
where $\Delta_{m,n}(t)=E_m(t)-E_n(t)$ and 
\begin{equation}\label{eps_mn}
\epsilon_{m,n}(t)=\frac{\langle E_m(t)|\dot{\hat{H}}(t)|E_n(t)\rangle}{\Delta^2_{m,n}(t)}\;
\end{equation}
are non-adiabatic parameters. We denote the typical scale of Eq.~(\ref{eps_mn}) as $\epsilon_{m,n}\sim\epsilon$, which is discussed more explicitly in Appendix~\ref{app:Uapprox} and serves as a small expansion parameter of the non-adiabatic perturbation theory. Writing the exact wavefunction as 
\begin{align}\label{psit}
|\psi (t)\rangle & = \hat{U}(t)|E_1(0) \rangle \nonumber \\
&= e^{i\gamma_1(t)}e^{-i\phi_1(t)} \sum_m c_m(t) |E_m(t)\rangle\;,
\end{align}
the amplitudes with $m \neq 1$ are given as follows, to leading order in $\epsilon$:
\begin{equation}\label{cm}
c_m(t)\simeq -i\epsilon_{m,1}(t)
+i \epsilon_{m,1}(0)e^{i\gamma_{m,1}(t)}e^{-i\phi_{m,1}(t)}\;,
\end{equation}
where  $\gamma_{m,n}(t)=\gamma_m(t)-\gamma_n(t)$ and $\phi_{m,n}(t)=\phi_m(t)-\phi_n(t)$. From the normalization of the state we obtain:
\begin{equation}
|c_1(t)| \simeq  1- \frac12 \sum_{m\neq 1} \left|\epsilon_{m,1}(t)-\epsilon_{m,1}(0)e^{i\gamma_{m,1}(t)}e^{-i\phi_{m,1}(t)}\right|^2,
\end{equation}
showing that, to linear order in $\epsilon$, only the phase of $c_1(t)$ is affected. Defining $c_1(t) = |c_1(t)|e^{i\delta\gamma_1(t)}$, the leading-order expression of $\delta\gamma_1(t)$ reads
\begin{equation}\label{dgamma}
\delta\gamma_1(t)\simeq \sum_{m\neq 1}\int_0^t dt' \Delta_{m,1}(t')\left|\epsilon_{m,1}(t')\right|^2.
\end{equation}
Note that, despite the formal dependence on $\sim \epsilon^2$, in the integrand, the integration domain of Eq.~(\ref{dgamma}) is $t \sim t_{\rm f}$, which diverges in the adiabatic limit. Therefore  Eq.~(\ref{dgamma}) can be estimated as $ \sim \epsilon^2 \Delta_{m,1} t_{\rm f} $, and it is not difficult to infer from Eq.~(\ref{eps_mn}) that $\Delta_{m,1} t_{\rm f} \sim \epsilon^{-1}$. Therefore, the phase correction is indeed of order $\epsilon$.

The above results are in agreement with previous literature \cite{Garrison1986, Rigolin2008, DeGrandi2010}. We refer the reader, in particular, to the detailed discussions of Ref.~\cite{Rigolin2008}. For completeness, in Appendix~\ref{app:Uapprox} we present an alternative derivation of Eqs.~(\ref{cm}) and (\ref{dgamma}), based on standard time-dependent perturbation theory. A main motivation of including Appendix~\ref{app:Uapprox} is to show how the textbook perturbative expansion can be made consistent with previous dervations. In particular, it was noted in Ref.~\cite{Rigolin2008} that, using the `standard approach', lowest-order perturbation theory will miss the phase correction given by Eq.~(\ref{dgamma}). Thus, a different type of non-adiabatic expansion was developed in that article. Instead, in Appendix~\ref{app:Uapprox} we show explicitly that the problem can be solved considering second-order perturbation theory. In fact, the perturbative order in terms of $\hat{H}_1$ and the order in the non-adiabatic expansion parameter $\epsilon$ do not necessarily coincide. While the $\sim \epsilon$ phase correction of Eq.~(\ref{dgamma}) is absent in first-order perturbation theory, it can be recovered by analyzing the second-order contribution (as reflected by the $\sim \epsilon^2$ dependence of the integrand). 

Adiabatic perturbation theory indicates that the fidelity can approach the ideal limit if we choose a sufficient smooth pulse at the initial and final times. More explicitly, setting in Eq.~(\ref{cm})
\begin{equation} \label{dGdt}
\dot{G}_j(0)=\dot G_j (t_{\rm f}) = 0,
\end{equation}
 we obtain
\begin{equation}\label{F_unitary}
F= \left| \langle E_1(t_{\rm f})| \psi(t_{\rm f}) \rangle\right|^2 \simeq  1,
\end{equation}
where the corrections are of order $\epsilon^2$. Formally, one can pursue further the non-adiabatic perturbation expansion and impose conditions analogous to Eq.~(\ref{dGdt}) to the higher-order derivatives of $G_j(t)$, obtaining an error which scales faster than $\epsilon^2$. In the following we will assume that the pulse is sufficiently smooth at $t=0$ and $t=t_{\rm f}$, such that the loss of fidelity induced directly by the non-adiabatic evolution becomes negliginle (with respect to the dissipative losses). Still, even for such a smooth time dependence, the non-adiabatic transitions are of considerable importance. The first term of Eq.~(\ref{cm}) cannot be zero for arbitrary $t$, implying that non-adiabatic transitions are always induced at intermediate times.  An adiabatic state transfer protocol is useful when $|E_1(t)\rangle$ is a `dark' state, i.e., it is especially robust to decoherence. However, the other instantaneous eigenstates do not necessarily share this property. Therefore, transitions to other states $|E_{m>1}(t)\rangle$ during the state transfer protocol make the system fragile to decoherence, and can induce a significant loss of fidelity.

\subsection{Effect of dissipation}\label{sec:fidelityTLME}
We now consider the effect of decoherence, which is modeled through a master equation of the following form:
\begin{equation}
    \label{eq:genericsystemL}
    \dot{\hat \rho}(t)=-i[\hat{H}(t),\hat\rho(t)]+\mathcal{R}(t) \hat\rho(t)\;,
\end{equation}
where $\mathcal{R}(t) \hat\rho  = \sum_\alpha\left( \hat{A}_\alpha(t)\hat\rho \hat{B}_\alpha(t)+\hat{C}_\alpha(t) \hat\rho+\hat\rho \hat{D}_\alpha(t)\right)$, for suitable time-dependent operators $\{\hat{A}_\alpha,\hat{B}_\alpha,\hat{C}_\alpha,\hat{D}_\alpha\}$. Such general form of a time-local master equation is also applicable to non-Markovian evolution, e.g., the Quantum Brownian motion \cite{Petruccione,Laine_2012,Garrahan2015,PhysRevA.31.1059,Andersson2014}, and recovers the standard Lindblad form as a special case. 
Here we are interested in an almost-ideal adiabatic state transfer process, characterized by a small loss of fidelity dominated by decoherence. Therefore, it is appropriate to consider the following solution of Eq.~(\ref{eq:genericsystemL}):
\begin{align}\label{rho_first_orger_Gamma}
\hat\rho(t) \simeq & \, |\psi (t) \rangle\langle\psi(t)|+ 
\int_0^t dt' \hat{U}(t)\hat{U}^{\dag}(t')\nonumber \\
& \times\bigg(\mathcal{R} (t')
 |\psi (t') \rangle\langle\psi(t')|\bigg)
\hat{U}(t')\hat{U}^{\dag}(t)\;,
\end{align}
where  $|\psi (t) \rangle$ is the exact unitary evolution, see Eq.~(\ref{psit}), and we have included the effect of decoherence to leading-order in $\mathcal{R}$. We then use this expression to compute the fidelity at the final time:
\begin{equation}\label{F_def}
F=\langle E_1(t_{\rm f}) |\hat\rho(t_{\rm f})|E_1(t_{\rm f}) \rangle.
\end{equation}
To simplify the above Eq.~(\ref{F_def}), we consider the smooth pulses discussed in the previous section, satisfying Eqs.~(\ref{dGdt}) and~(\ref{F_unitary}). In that case we can write $|\psi(t_{\rm f})\rangle= \hat{U}(t_{\rm f})|E_1(0)\rangle \simeq e^{-i\varphi_1}|E_1(t_{\rm f})\rangle$, where $\varphi_1 \simeq \phi_1(t_{\rm f})-\gamma_1(t_{\rm f})-\delta\gamma_1(t_{\rm f})$. From this, it also follows that:
\begin{equation}\label{Udag_approx}
\hat{U}(t')\hat{U}^\dag(t_{\rm f}) |E_1(t_{\rm f})\rangle\simeq  e^{i \varphi_1}  |\psi(t')\rangle .
\end{equation}
Making use of these approximations, we can finally write Eq.~(\ref{F_def}) in the following form:
\begin{align} \label{eq:genericfidelityloss}
F\simeq  1 &  + \sum_\alpha \int_0^{t_{\rm f}} dt  \bigg(
\langle\psi(t)|\hat{C}_\alpha(t)+\hat{D}_\alpha(t)|\psi(t)\rangle
 \nonumber \\
& + \langle\psi(t)|\hat{A}_\alpha(t)|\psi(t)\rangle\langle\psi(t)|\hat{B}_\alpha(t)|\psi(t)\rangle\bigg),
\end{align}
which is a main result of this section and, combined to Eq.~(\ref{cm}) for the expansion coefficients of $|\psi(t)\rangle$, forms the basis of our following discussions.

\subsection{Fidelity loss in terms of a Lagrangian}\label{Fas_Lagran}

To leading order in non-adiabatic perturbation theory, $|\psi(t)\rangle$ only depends on the control parameters ${\bf G}$ and their derivatives, which motivates rewriting Eq.~(\ref{eq:genericfidelityloss}) in the form of a classical action:
\begin{equation}\label{DeltaF}
\Delta F = 1- F \simeq  \int_0^{t_{\rm f}} L({\bf G},\dot{\bf G},t)dt .
\end{equation}
To be more specific, we will discuss from now on a time-independent $\mathcal{R}$ of the Lindblad form, i.e., we take $\hat{B}_\alpha = \hat{A}_\alpha^\dag$ and $\hat{C}_\alpha = \hat{D}_\alpha =-\frac12 \hat{A}_\alpha^\dag\hat{A}_\alpha$. Then, the classical Lagrangian can be derived from the general expression:
\begin{align} \label{eq:fidelitylossLindblad}
L = \sum_\alpha \bigg(  \langle  \psi(t) |
 \hat{A}_\alpha^\dag\hat{A}_\alpha|\psi(t)\rangle
- \left|\langle\psi(t)|\hat{A}_\alpha|\psi(t)\rangle \right|^2 \bigg).
\end{align}
As we will see shortly, we can further separate in $L$ the contribution of dissipative processes causing a direct loss of fidelity from the terms which combine dissipation and non-adiabatic leakage. This distinction allows us to identify the potential and kinetic terms of the Lagrangian, $L = K - V$.  

We first consider the derivation of the effective potential, obtained by applying to Eq.~(\ref{eq:fidelitylossLindblad}) the adiabatic approximation, $|\psi(t)\rangle \simeq e^{i\gamma_1(t)-i\phi_1(t)}|E_1(t)\rangle$. This gives: 
\begin{align} \label{eq:Vpotential}
V({\bf G}) = - {\sum_\alpha} \bigg(  \langle  E_1 |
 \hat{A}_\alpha^\dag\hat{A}_\alpha|E_1 \rangle
- \left|\langle E_1|\hat{A}_\alpha|E_1\rangle \right|^2 \bigg).
\end{align}
In Eq.~(\ref{eq:Vpotential}) we have omitted the time dependence of $|E_1 (t)\rangle$, which is only induced through the couplings $G_j(t)$. For a Hamiltonain as in Eq.~(\ref{eq:genericH}), the instantaneous eigenstates are determined by the generalized coordinates ${\bf G}$, giving rise to a potential $V({\bf G})$ which has no explicit time dependence. 

We now consider dissipative processes which do not contribute to Eq.~(\ref{eq:Vpotential}), thus the non-adiabatic corrections should be considered explicitly. Suppose that for a given $\alpha$ we have 
$\langle  E_1(t) | \hat{A}_\alpha^\dag\hat{A}_\alpha|E_1(t)\rangle - |\langle  E_1(t) |
 \hat{A}_\alpha|E_1(t)\rangle |^2 =  0$, i.e., process $\alpha$ does not contribute to $V({\bf G})$. By inserting a completeness relation with respect to $|E_m(t)\rangle$, this condition can also be expressed as:
  \begin{equation}\label{dark_states_condtition}
 \langle E_{m\neq 1}(t) | \hat{A}_\alpha | E_1(t) \rangle = 0.
 \end{equation}
Making use of Eqs.~(\ref{psit}) and (\ref{dark_states_condtition}) in Eq.~(\ref{eq:fidelitylossLindblad}), and neglecting $o(\epsilon^2)$ terms, we obtain:
\begin{align}\label{K_final_expession}
 K \simeq & {\sum_\alpha}^\prime \sum_{l>1} 
\Big| c_l \langle E_1|\hat{A}_\alpha|E_1 \rangle - \sum_{m>1}c_m \langle  E_l |\hat{A}_\alpha|E_m \rangle  \Big|^2,
 \end{align}
 where the summation over $\alpha$ is restricted to values which satisfy Eq.~(\ref{dark_states_condtition}). For convenience of the reader, we present in Appendix~\ref{positive_kinetic} detailed steps leading to Eq.~(\ref{K_final_expession}). The analogy to a kinetic term becomes clear when considering the expression of $c_n$ from non-adiabatic perturbation theory, given in Eq.~(\ref{cm}). Due to Eq.~(\ref{dGdt}), we simply have $c_m(t) \simeq -i \epsilon_{m,1}(t)$ which, using Eqs.~(\ref{eps_mn}) and (\ref{eq:genericH}), gives: 
 \begin{equation}\label{cm_Gdot}
c_n \simeq  -i \sum_j \dot{G}_j \frac{\langle E_n|\hat{H}_j|E_1\rangle}{(E_n -E_1)^2}.
\end{equation}
Therefore, $K$ is a quadratic form of the $\dot{G}_j$:
 \begin{equation}\label{Mtensor}
 K({\bf G},\dot{\bf G}) = \frac12 \sum_{ij} [M^{-1}({\bf G})]_{ij}\dot{G}_i\dot{G}_j,
 \end{equation}
 where $M^{-1}$ plays the role of an inverse effective mass tensor. Its explicit expression is found by inserting Eq.~(\ref{cm_Gdot}) into Eq.~(\ref{K_final_expession}). Like $V({\bf G})$, the mass tensor only depends on the generalized coordinates ${\bf G}$. As it is natural to expect, the `potential' and `kinetic' contributions should lead to a loss of fidelity. This can be easily checked from the explicit expressions: The potential in Eq.~(\ref{eq:Vpotential}) is always negative, while the kinetic term in Eq.~(\ref{K_final_expession}) is obviously positive.

After having presented the above derivations, it is interesting to add a few  comments on the order of approximation of different terms. In fact, the $V$ term is obtained through the adiabatic approximation while the $K$ term arises from non-adiabatic corrections. Therefore, one could expect the $K$ term to to be highly suppressed, thus of little significance for the fidelity loss. This is, however, not the case. The apparent inconsistency in tracking the various orders of approximation is related to the typical scale of the decay rate $\Gamma_\alpha$  associated with each type of processes ($\hat{A}_\alpha \propto \sqrt{\Gamma_\alpha}$). Since the adiabatic state transfer is designed to evolve $|E_1(t)\rangle$ within a `dark' subpspace, which has exceptional coherence properties, the rates typically satisfy $\Gamma_{\alpha_V} \ll \Gamma_{\alpha_K}$, where $\alpha_V$ ($\alpha_K$) labels a process contributing to the direct (indirect) fidelity loss. Due to their large decoherence rate, processes contributing to $K$, which are suppressed by adiabaiticty, can have an effect comparable to the ones acting within the `dark' subspace. 
On the other hand, it is justified to omit non-adiabatic effects for the processes of the first type, i.e., contributing to $V$. These corrections are of higher-order with respect to Eq.~(\ref{eq:Vpotential}), thus they can be safely neglected.
 
If we wish to maximize the transfer fidelity, we expect the `potential' and `kinetic' contributions to play an equally important role. Supposing that  Eq.~(\ref{eq:Vpotential}) causes a large fidelity loss, $\Delta F$ can be reduced in a natural way by decreasing $t_{\rm f}$. In this way, dissipation acts on the `dark' state for a shorter time. However,doing so will generally enhance the contribution of Eq.~(\ref{K_final_expession}). In the opposite scenario of a larger `kinietic' fidelity loss, it becomes advantageous to suppress the non-adiabatic processes by a longer transfer time $t_{\rm f}$, thus increasing the effect of disipation acting on the `dark' subspace. In general, as we stated already, both terms should be accounted for on equal footing to achieve an optimal state transfer, which requires a careful balance of direct decoherence and non-adiabatic dissipative effects \cite{wang2017optimization}.

 In conclusion, we have cast the loss of fidelity of an adiabatic state-transfer process in the form of a classical action  \cite{CMechanics}. For a given transfer time $t_{\rm f}$, the optimal path in parameter space can be obtained from the variational condition $\delta \Delta F = 0$, i.e., by applying the principle of least action. The corresponding Euler-Lagrange equations are:
\begin{equation}\label{eq:E-L}
     \frac{d}{dt}\left(\frac{\partial L({\bf G},\dot{\bf G})}{\partial { \dot{G}_j}}\right)=\frac{\partial L({\bf G},\dot{\bf G})}{\partial G_j},
 \end{equation}
where $L({\bf G},\dot{\bf G}) = K({\bf G},\dot{\bf G})-V({\bf G})$ is given by Eqs.~(\ref{eq:Vpotential}), (\ref{K_final_expession}), and (\ref{cm_Gdot}). From the optimal path ${\bf G} (t)$, the transfer fidelity is directly obtained from the action, see Eq.~(\ref{DeltaF}). Finally, after performing the functional minimization at given $t_{\rm f}$, it is in principle quite straighforward to determine the optimal transfer time.

In the following section we will apply the above formalism to the state transfer between two qubits coupled through a highly dissipative quantum bus, obtaining a general characterization of the optimal fidelity.

\section{Application: indirect state transfer between two qubits}\label{Model and perturbation}
We consider here a model of two qubits resonantly coupled to a bosonic quantum bus. Using the rotating wave approxiamtion, the Hamiltonian can be written as follows
\begin{equation}\label{eq:LambdaH}
    \hat{H}(t)=\omega a^\dagger a+\sum_{j=1,2}\frac{\omega}{2}\hat{\sigma}_j^z+ G_j(t)\left(\hat a^\dagger\hat\sigma_j^-+\hat{a}\hat\sigma_j^+\right),
\end{equation}
where the quantum bus is described by a single harmonic mode with annihilation operator $\hat{a}$, $\hat{\sigma}_j^\pm = (\hat{\sigma}_j^x \pm i \hat{\sigma}_j^y)/2$, where $\hat{\sigma}_j^\alpha$ ($\alpha = x,y,z$) are the Pauli matrices of qubit $j$, and $G_j(t)$ are controllable couplings between the quantum bus and the $j$-th qubit, which we take as real. The model dynamics is given by Eq.~(\ref{eq:genericsystemL}), where we consider the following dissipative term:
\begin{align}\label{dissipation}
\mathcal{R} \hat\rho=&\sum_{j=1,2}\left(\gamma_j^R D[\hat\sigma_j^-]\hat\rho+\gamma_j^\phi D[\hat\sigma_j^z]\hat\rho\right) \nonumber \\
&+\kappa^R D[\hat a]\hat\rho+\kappa^\phi D[\hat a^\dagger \hat a]\hat\rho\;,
\end{align}
which is a quite general model of a low-temperature Markovian environment. In Eq.~(\ref{dissipation}), $D[\hat A]\hat\rho=\hat A\rho \hat A^\dagger-\frac12\hat A^\dagger \hat A\rho-\frac12\hat\rho \hat A^\dagger \hat A$ and $\gamma_j^R,\gamma_j^\phi$ are, respectively, the decay and pure dephasing rate  of the $j$-th qubit, while $\kappa^R, \kappa^\phi$ are the analogous rates of the bosonic mode.  For an highly dissipative quantum bus, i.e., $\kappa^R, \kappa^\phi \gg \gamma_j^R,\gamma_j^\phi$, it becomes advantageous to perform a state transfer between the qubits through an adiabatic protocol.

\begin{figure}
\centering
\includegraphics[width=0.8\columnwidth]{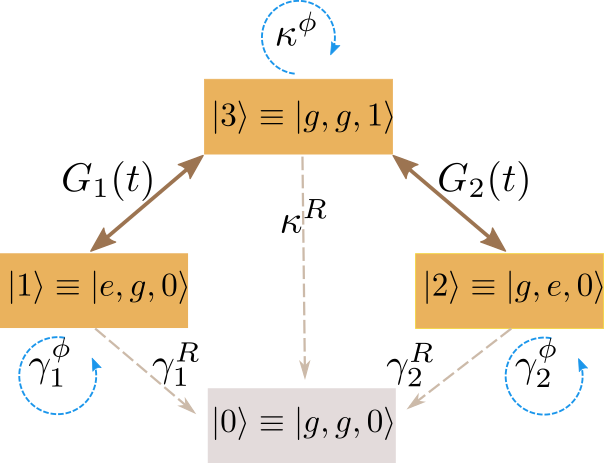}
\caption{Schematics of relevant states and processes involved in the state transfer $|1\rangle \to |2\rangle$.  In the low-temperature limit $k_B T\ll\hbar \omega$, the system can be restricted to a 4-level subspace. The thick double arrows represent the coherent couplings  of Eq.~(\ref{eq:LambdaH}). Dashed arrows show the relaxation (brown) and dephasing (blue) pathways of Eq.~(\ref{dissipation}).\label{sys}}
\end{figure}

The relevant couplings and dissipative mechanisms are summarized in Fig.~\ref{sys}. Due to the approximation of a zero temperature bath, the model dynamics can be restricted to the four-level subspace spanned by $\mathcal{S}=\{|e, g,0\rangle, |g, e, 0\rangle, |g, g, 1\rangle,|g, g, 0\rangle \}$~\cite{wang2017optimization}, where $|e\rangle$ and $|1\rangle$ ($|g\rangle$ and $|0\rangle$) denote the excited (ground) state of the qubits and the bus, respectively. The states with one excitation form a $\Lambda$-system, whose instantaneous eigenstates are easily found. We thus have:
\begin{align}\label{eigenstates}
&|E_0\rangle = |g,g,0\rangle, \nonumber \\
&|E_1\rangle =\cos\theta|e,g,0\rangle -\sin\theta|g,e,0\rangle ,\nonumber \\
&|E_\pm \rangle = \frac{1}{\sqrt{2}}(\sin\theta|e,g,0\rangle +\cos\theta|g,e,0\rangle \pm |g,g,1\rangle),
\end{align}
with energies $E_0 = E_1 = 0$ and $E_\pm = \pm  G$. Here we have defined the important parameters:
\begin{equation}
\theta=\arctan\frac{G_1}{G_2}, ~~~~~ G=\sqrt{G_1^2+G_2^2}.
\end{equation}

From Eq.~(\ref{eigenstates}) we easily see that the adiabatic evolution of $\alpha |E_0\rangle + \beta |E_1\rangle $ with the boundary conditions:
\begin{equation}
    \label{eq:thetabd}
    \theta(0)=0, \qquad \theta(t_{\rm f})=\frac{\pi}{2}\;,
\end{equation}
will transfer a generic state from qubit~$1$ qubit~2. More precisely, the state $\alpha|g\rangle + \beta |e\rangle$ of qubit 1 is transferred to $\alpha|g\rangle - \beta |e\rangle$ of qubit 2. Since $|E_0\rangle,|E_1\rangle$ are dark states, which do not excite the quantum bus, the adiabatic limit realizes an ideal state transfer in the absence of qubit decoherence. In the realistic case with $\gamma_i^{R,\phi} \neq 0$ we set below $\beta=1$, i.e., consider the state transfer of the excited state $|e\rangle$. For this model, state $|e\rangle$ typically limits the overall fidelity, since $|E_0\rangle$ is not affected by the time evolution (thus, the state transfer of $|g\rangle$ is ideal). 

By applying to state $|e\rangle$ the formulas of the previous section, we find the transfer fidelity loss as follows:
\begin{align}\label{Delta_F_Lambda}
\Delta F =\int_0^{t_{\rm f}} dt \Bigg[ & \kappa_{\rm tot}\frac{\dot\theta(t)^2}{G(t)^2}+ \gamma^\phi_{\rm tot} \sin^2 2\theta(t)  \nonumber \\
& + \gamma_1^R \cos^2\theta(t) + \gamma_2^R \sin^2 \theta(t) \Bigg],
\end{align}
where $\kappa_{\rm tot} = \kappa^R + \kappa^\phi$ and $\gamma^\phi_{\rm tot} = \gamma^\phi_1 + \gamma^\phi_2$. Since the first term in the square parenthesis is the only one containing $G$, the fidelity  is obviously improved at larger coupling strength. $G$, however, is subject to physical constrains and we indicate with $G_{\rm max}$ the largest achievable value.

\begin{figure}
\centering
\includegraphics[width=0.4
\textwidth]{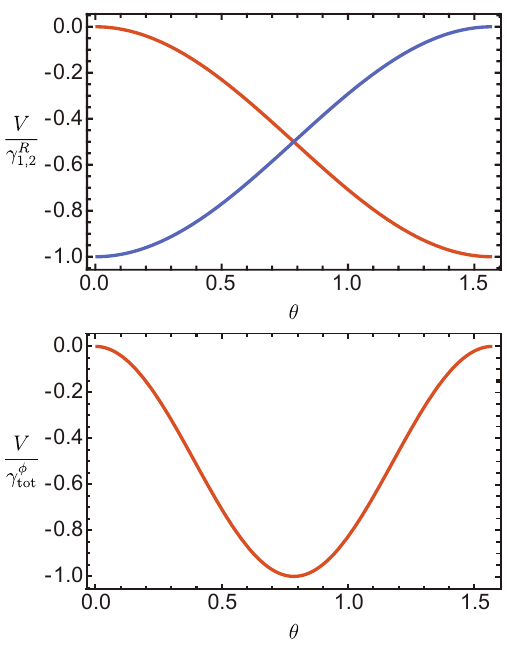}
\caption{Limiting forms of $V(\theta)$. In the upper panel there is only relaxation of the first (blue curve) or second qubit (orange curve). The lower panel is for pure dephasing.\label{pic:V}   }
\end{figure}

In several relevant scenarios we can assume that the maximum total coupling is a function of $\theta$:
\begin{equation}\label{Gmax_general}
G(t) = G_{\rm max}(\theta(t)).
\end{equation}
The simple choice of a constant $G_{\rm max}$ gives a parallel adiabatic passage (PAP), where the instantaneous eigenvalues $E_\pm$ do not change with time. Following Ref.~\cite{wang2017optimization}, we will also consider a more concrete model for $G_{\rm max}$ where system coupling $G_{1/2}$ is bounded by $G_{1/2,{\rm max}}$. Then, for a given value of $\theta$, we find the following expression for the maximum coupling strength:
\begin{equation}\label{eq:Gmax}
     G_{\rm max}(\theta)=\begin{cases} G_{2,\mathrm{max}}/\cos\theta, &\theta\in(0,\bar{\theta})\\
    G_{1,\mathrm{max}}/\sin\theta,  &\theta\in(\bar{\theta},\pi/2)
    \end{cases}
\end{equation}
with $\bar{\theta}=\arctan G_{1,\rm max}/G_{2,\rm max}$. Adopting Eq.~(\ref{Gmax_general}), the Lagrangian becomes time-independent. It describes a 1D particle with coordinate $\theta \in [0,\pi/2]$, effective mass $G_{\rm max}^2/2\kappa_{\rm tot}$ (which, in general, depends on $\theta$), and the potential:
\begin{equation}
V(\theta) =  - \left( \gamma_1^R \cos^2\theta + \gamma_2^R \sin^2 \theta + \gamma^\phi_{\rm tot}\sin^2 2\theta \right),
\end{equation}
whose limiting cases are represented in Fig.~\ref{pic:V}. As shown by the blue curve of the upper panel, relaxation of the first qubit gives a potential barrier. The particle has a large initial velocity at $\theta=0$ and slows down at the top of the barrier, i.e., the excitation is quickly moved away from the first qubit and the system spends more time close to $\theta=\pi/2$, when the state $|e\rangle$ has been nearly transferred to the second (and more coherent) qubit. An opposite behavior occurs when there is only relaxation of the second qubit (orange curve of the upper panel). Finally, if dephasing is large, the potential tends to accelerate before reaching the minimum at $\theta = \pi/4$, when the excitation is coherently shared by both qubits. In general, the competition between different decoherence mechanisms will result in a more complex potential profile.

\subsection{Fidelity Optimization}\label{optimization}

The optimization is most easily performed using that the energy $\mathcal{E}$ of the particle is conserved, which gives:
\begin{equation}\label{dtheta}
\dot{\theta}= G_{\rm max}(\theta)\sqrt{\frac{\mathcal{E}-V(\theta)}{\kappa_{\rm tot}}}.
\end{equation}
Note that the above expression does not automatically satisfy the boundary condition in Eq.~(\ref{dGdt}), which implies $\dot\theta(0)=\dot{\theta}(t_{\rm f})=0$. For a more in-depth discussion of this apparent inconsistency,  see Appendix~\ref{app:boundaryc}. From the above Eq.~(\ref{dtheta}), the time dependence $\theta(t)$ can be easily obtained.  In our context, it is useful to consider $\mathcal{E}$ as a parameterization of the transfer time $t_{\rm f} = \int_0^{t_{\rm f}} d\theta/\dot\theta$. Explicitly, we have:
\begin{equation}\label{tf_from_E}
t_{\rm f} = \int_0^{\pi/2} \frac{d\theta}{G_{\rm max}(\theta)}\sqrt{\frac{\kappa_{\rm tot}}{\mathcal{E}-V(\theta)}}.
\end{equation}
Meanwhile, we can express the optimal fidelity as:
\begin{equation}
\label{Fopt}
\Delta F_{\rm opt}=\int_0^{\pi/2}\frac{\sqrt{\kappa_{\rm tot}}}{G_{\rm max}(\theta)}\frac{\mathcal{E}-2V(\theta)}{\sqrt{\mathcal{E}-V(\theta)}}d\theta.
\end{equation}
We can further impose $\partial \Delta F_{\rm opt}/\partial \mathcal{E} =0 $, finding that the smallest error is at  $\mathcal{E} =0 $: 
\begin{equation}
\label{Fmin}
\Delta F_{\rm min}=2\int_0^{\pi/2}\frac{\sqrt{\kappa_{\rm tot}|V(\theta)|}}{G_{\rm max}(\theta)}d\theta.
\end{equation}
The minimization of Eq.~(\ref{Fopt}) with respect to $\mathcal{E}$ is equivalent to minimizing $\Delta F_{\rm opt}$ with respect to the transfer time. Thus, the optimal transfer time is simply obtained by setting $\mathcal{E} =0 $ in Eq.~(\ref{tf_from_E}). Form these optimization procedures, we can find the elementary variable is not time but $\theta$, the relation between the two couplings, which has fixed boundary values: $0, \pi/2$.

\begin{figure}
\centering
\includegraphics[width=0.3\textwidth]{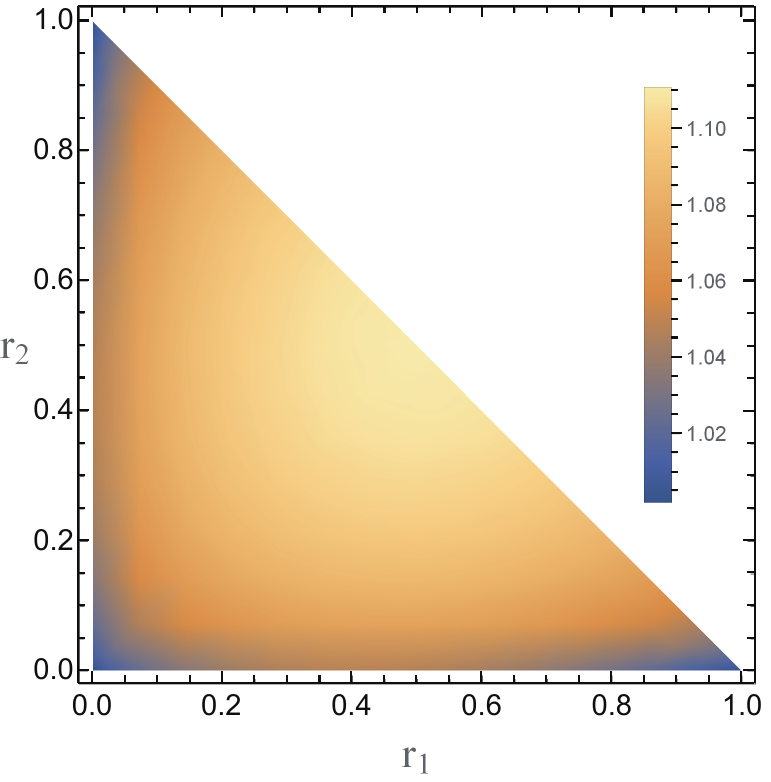}
\caption{The prefactor $c(r_1,r_2)$, appearing in Eq.~(\ref{FminPAP}). 
\label{c_plot} }
\end{figure}
To evaluate the above formulas, we need a specific form for the total coupling. As mentioned earlier, a simple choice is to take $G_{\rm max}$ as a constant value. In this case, we find:
\begin{equation}
\label{FminPAP}
\Delta F_{\rm min}^{\rm PAP}=2 c(r_1,r_2) \frac{\sqrt{\kappa_{\rm tot} \gamma_{\rm tot}}}{G_{\rm max}},
\end{equation}
where $\gamma_{\rm tot} = \gamma_1^R+\gamma_2^R+\gamma^\phi_{\rm tot}$ and $r_{1,2}=\gamma_{1,2}^R/\gamma_{\rm tot} $. The prefactor $c(r_1,r_2)$ is given by:
\begin{align}\label{cPAP}
c(r_1,r_2)=\int_0^{\pi/2}d\theta &\left[ r_1 \cos^2{\theta}+r_2 \sin^2{\theta} \right. \nonumber \\
&\left.+(1-r_1-r_2) \sin^2{2\theta}\right]^{1/2},
\end{align}
and can be easily obtained by numerical integration. As shown in Fig.~\ref{c_plot}, $c(r_1,r_2)$ has a weak dependence on $r_{1,2}$, and it is always close to 1. The allowed interval is:
\begin{equation}
1 \leq c(r_1,r_2) \leq \frac{\pi\sqrt{2}}{4},
\end{equation}
where the upper bound  $\pi\sqrt{2}/4 \simeq 1.11$. This value is obtained at $r_1=r_2=1/2$, i.e., with identical qubits and $\gamma_{\rm tot}^\phi=0$. It recovers the optimal fidelity computed in Ref.~\cite{wang2017optimization}, where dephasing was not considered. The lower bound is realized in the three extreme cases of Fig.~\ref{pic:V}. For example, we find  $c=1$  when there is dephasing but no  relaxation ($r_1=r_2=0$).

If, instead, we consider the more realistic case of Eq.~(\ref{eq:Gmax}), the following expression is found:
\begin{equation}
\label{Fmin2}
\Delta F_{\rm min}=  
2c(r_1,r_2,\bar{\theta})\sqrt{\frac{\kappa_{\rm tot} \gamma_{\rm tot}}
{G_{1,\rm max}^2+G_{2,\rm max}^2}},
\end{equation}
where the coefficient is given by:
\begin{align}\label{c_coeff_intgral}
 c(r_1,r_2,\bar{\theta}) = & 
\frac{1}{\sin\bar\theta} \int_0^{\cos\bar{\theta}} dx \sqrt{g(r_1,r_2,x)}  \nonumber \\
 & +\frac{1}{\cos\bar{\theta}} \int_0^{\sin\bar{\theta}} dx \sqrt{g(r_2,r_1,x)},
\end{align}
with $g(r_1,r_2,x)=r_1 x^2+r_2 (1-x^2)+4(1-r_1-r_2)x^2(1-x^2)$. We recall here that  $\bar{\theta}=\arctan G_{1,\rm max}/G_{2,\rm max}$ and notice that  the following relation holds:
\begin{equation}\label{c_transf}
c(r_1,r_2,\bar{\theta})=c(r_2,r_1,\pi/2 - \bar{\theta}).
\end{equation}
Physically, Eq.~(\ref{c_transf}) implies that the two qubits can be interchanged without affecting the optimal fidelity. In other words, the transfer fidelity has the same optimal value in both directions.


For several limiting cases, Eq.~(\ref{c_coeff_intgral}) can be integrated exactly. In particular, for identical relaxation rates and no dephasing ($\gamma_1^{R}=\gamma_2^{R}$ and $\gamma_{\rm tot}^\phi = 0 $) we recover the expression of Ref.~\cite{wang2017optimization}:
\begin{equation}
\label{FminPAP_case1}
\Delta F_{\rm min}=2 \sqrt{\kappa_{\rm tot}\gamma_1^R} \sqrt{\frac{1}{G_{1,\rm max}^2}+\frac{1}{G_{2,\rm max}^2}}.
\end{equation}
If, instead, relaxation of the first qubit is dominant ($\gamma_2^{R}=\gamma_{\rm tot}^\phi = 0 $), Eq.~(\ref{FminPAP}) gives:
\begin{equation}
\label{FminPAP_case2}
\Delta F_{\rm min}=\sqrt{\kappa_{\rm tot}\gamma_1^R}\left( \frac{1}{G_{1,\rm max}}+\frac{\bar\theta}{G_{2,\rm max}}
\right).
\end{equation}
If there is only relaxation of the second qubit ($\gamma_1^{R}=\gamma_{\rm tot}^\phi = 0 $), an expression analogous to  Eq.~(\ref{FminPAP_case2}) is found, with $1 \leftrightarrow 2$ (which also implies $\bar\theta \to \pi/2 - \bar\theta$). Finally, if there is only dephasing ($\gamma_1^{R}=\gamma_{2}^R = 0 $) we find:
\begin{equation}
\label{FminPAP_case3}
\begin{split}
\Delta F_{\rm min}=&\frac43 
\sqrt{\frac{\kappa_{\rm tot}\gamma_{\rm tot}^\phi}{G_{1,\rm max}^2+G_{2,\rm max}^2}}
\left( 
\frac{1}{\sin\bar\theta}+
\frac{1}{\cos\bar\theta}-1
\right)
\;.
\end{split}
\end{equation}

\begin{figure}
 \includegraphics[width=0.3\textwidth]{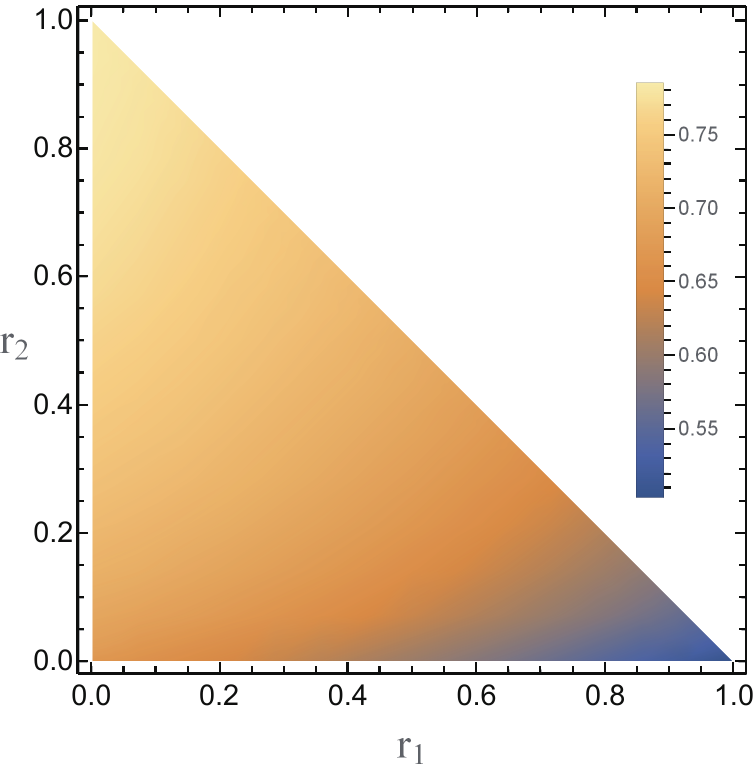}\qquad
\includegraphics[width=0.3\textwidth]{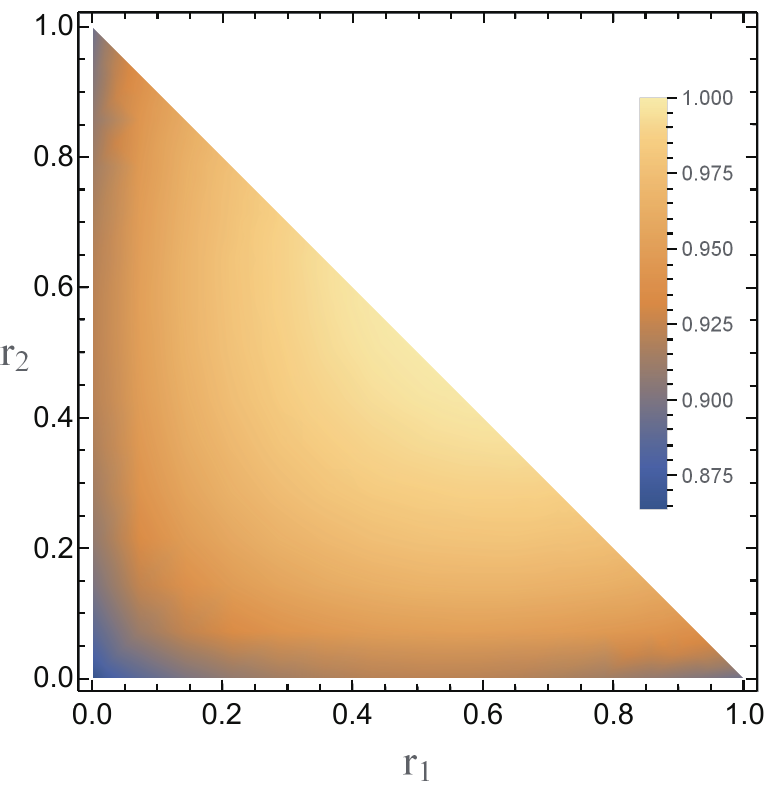}      
\caption{The prefactor $c_1(r_1,r_2,\bar{\theta})$, appearing in Eq.~(\ref{Fmin2_G1}). We take $\bar\theta =0$ and $\bar\theta =\pi/4$ in the upper and lower panel, respectively. Note that the two panels have different color scales (insets).\label{c1_plot} } 
\end{figure}

In the general case, the value of $c(r_1,r_2,\bar\theta)$ is easily found by numerical integration. However, Eq.~(\ref{Fmin2}) is written in a symmetric form with respect to the two qubits, which can become slightly misleading when one of the couplings $G_{i,\rm max}$ is much stronger than the other. For example, if in Eq.~(\ref{Fmin2}) we consider the limit $G_{2,\rm max} \to \infty$, one can easily see that the coefficient $c(r_1,r_2,\bar\theta \to 0)$ diverges, compensating the vanishing dependence of $G_{\rm max}$. The optimal fidelity tends to a well-defined limit, which only depends on $G_{1,\rm max}$. Then, to always have a prefactor of order unity, we express the optimal fidelity in terms of the weakest coupling, which is actually limiting the transfer fidelity. By restricting ourselves to $G_{1,\rm max} \leq G_{2,\rm max}$ ($\bar{\theta}\leq\pi/4$), we simply rewrite Eq.~(\ref{Fmin2}) as:
\begin{equation}
\label{Fmin2_G1}
\Delta F_{\rm min}=  
2 c_1(r_1,r_2,\bar{\theta})
\frac{\sqrt{\kappa_{\rm tot} \gamma_{\rm tot}}}{G_{1,\rm max}},
\end{equation}
where $c_1(r_1,r_2,\bar{\theta})= c(r_1,r_2,\bar{\theta})\sin\bar\theta$. The dependence of $c_1$ on $r_{1,2}$ is shown in Fig.~\ref{c1_plot} for the two limiting cases $G_{1,\rm max}\ll G_{2,\rm max}$ (upper panel) and $G_{1,\rm max}=G_{2,\rm max}$ (lower panel). We see that the value of the coefficient $c_1$ lies in the range:
\begin{equation}
    \frac12 \leq c_1(r_1,r_2,\bar\theta) \leq 1
\end{equation}
where the lower bound is $c_1 (1,0,0)$ (see the upper panel of Fig.~\ref{c1_plot}) and the upper bound is found $c_1 (1/2,1/2,\pi/4)$ (see the lower panel of Fig.~\ref{c1_plot}). In general,  as seen in Fig.~\ref{c1_plot}, the case  $G_{1,\rm max}\ll G_{2,\rm max}$ yields smaller values of $c_1$, thus is more favorable for obtaining a higher transfer fidelity. However, the comparison between the two panels also shows that, at a given value of $G_{1,\rm max}$, only a modest gain of fidelity is obtained by a large increase of $G_{2,\rm max}$. In general, we conclude from Eq.~(\ref{Fmin2_G1}) that the loss of fidelity is of order $\Delta F_{\rm min} \sim
\sqrt{\kappa_{\rm tot} \gamma_{\rm tot}}/G_{1,\rm max}$ or, more generally:
\begin{equation}
\Delta F_{\rm min} \sim
\frac{\sqrt{\kappa_{\rm tot} \gamma_{\rm tot}}}{{\rm min}\{G_{1,\rm max},G_{2,\rm max}\}}\;,
\end{equation}
which is compatible with the limiting cases Eqs.~(\ref{FminPAP_case1}--\ref{FminPAP_case3}).

We finally consider the optimal transfer time, which for the specific case of Eq. ~(\ref{FminPAP_case1}), i.e.,  $\gamma_1^{R}=\gamma_2^{R}$ and $\gamma_{\rm tot}^\phi = 0 $, is found as:
\begin{equation}
\label{FminPAP_tmin}
 t_{\rm f}^*= \sqrt{\frac{2\kappa_{\rm tot}}{\gamma_1^R} \left(\frac{1}{G_{1,\rm max}^2}+\frac{1}{G_{2,\rm max}^2}\right)}.
\end{equation}
This expression suggests that:
\begin{equation}
\label{FminPAP_tmin_estimate}
t_{\rm f}^* \sim \frac{\sqrt{\kappa_{\rm tot}/\gamma_{\rm tot}}}{{\rm min}\{G_{1,\rm max},G_{2,\rm max}\}}\;.
\end{equation}
Unfortunately, however, the estimate of the optimal transfer time is more subtle than Eq.~(\ref{FminPAP_tmin_estimate}). For example, in the limiting case of Eq.~(\ref{FminPAP_case2}), where $\gamma_2^{R}=\gamma_{\rm tot}^\phi = 0 $, the optimal transfer is found to diverge, $t^*_{\rm f} \to \infty$.  The reason is actually quite simple: Since in this case the second qubit does not suffer dissipation, the system can spend a long time close to the final state ($\theta=\pi/2$), without compromising the transfer fidelity. A similar situation occurs in the case of pure dephasing, see Eq.~(\ref{FminPAP_case3}). In general, however, we expect Eq.~(\ref{FminPAP_tmin_estimate}) to hold in the reasonable scenario of two qubits with comparable decoherence rates. For two qubits of the same type, we can also reasonably expect $G_{1,\rm max} \sim G_{2,\rm max}  \sim G_{\max} $.

For similar qubits, the estimate in Eq.~(\ref{FminPAP_tmin_estimate}) gives an optimal transfer time $t_{\rm f}^* \gg 1/G_{\rm max}$ (taking into account that $\kappa_{\rm tot} \gg \gamma_{\rm tot}$). This condition is compatible with the quasi-adiabatic evolution, showing that our optimization is internally consistent. Instead, if we take Eq.~(\ref{FminPAP_tmin_estimate}) literally and apply it to the the regime $\kappa_{\rm tot}\sim\gamma_{\rm tot}$, it gives $t_{\rm f}^* \sim 1/G_{\rm max} $. This suggests that, when the dissipation rates are comparable, the transfer should be executed as fast as possible  and is limited by the largest achievable coupling strength. The corresponding fidelity loss is $\Delta F \sim \gamma_{\rm tot}/G_{\rm max}$. While these conclusions are reasonable, our formalism is clearly not applicable to this regime, and other types of optimization schemes should be considered.

\begin{figure}
\centering
\includegraphics[width=0.45\textwidth]{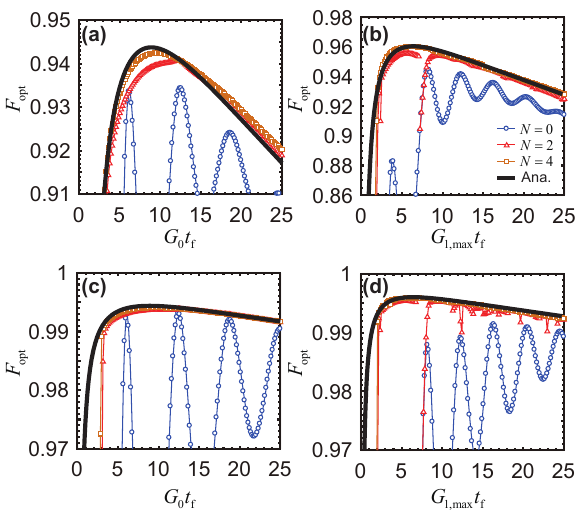}
\caption{Optimal fidelity $F_\mathrm{opt}\equiv1-\Delta F_\mathrm{opt}$ as a function of state transfer time $t_\mathrm{f}$. In each panel, the black solid line is computed from the least-action principle, while the markers along the other three curves are obtained by direct numerical optimization, based on Eq.~(\ref{FourierExpansion}) with $N=0,2,4$ (from bottom to top). Panel (a) is for parallel adiabatic passage (PAP), with  $\kappa^R = 0.1$, $\gamma^R_1=2.5\times 10^{-3}$, $\gamma^R_2=2\times 10^{-3}$,  and $\gamma^\phi_1=\gamma^\phi_2 = 10^{-3}$ (setting $G_{\rm max}=1$). In panel (b) the restriction on the couplings is given by Eq.~(\ref{eq:Gmax}), with $G_{2,\rm max} =2$ (setting $G_{1,\rm max} =1$). In these units, all dissipation rates are the same of panel (a). Panels (c) and (d) are like panels (a) and (b), respectively, except that all dissipative rates are ten times smaller: $\kappa^R = 0.01$, $\gamma^R_1=2.5\times 10^{-4}$, $\gamma^R_2=2\times 10^{-4}$,  and $\gamma^\phi_1=\gamma^\phi_2 = 10^{-4}$. In all panels, $\kappa^\phi = 0$. }
\label{Numerics}
\end{figure}

\begin{figure}
\begin{centering}
\includegraphics[width=0.45\textwidth]{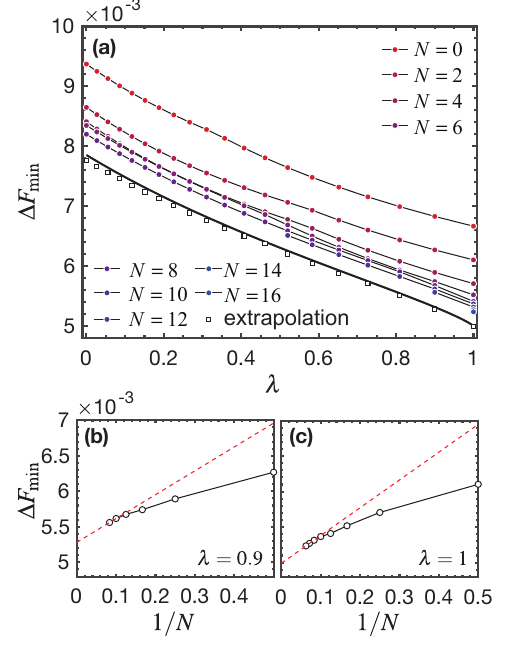}
\par\end{centering}
\caption{
Minimum fidelity loss, optimized over the transfer time $t_{\rm f}$ and extrapolated to an infinite number of Fourier coefficients $\alpha_n$. Panel (a) shows, for a PAP process, numerically optimized values of $\Delta F_\mathrm{min}$ (colored circles), obtained at different values of $\lambda=(\gamma_2^R-\gamma_1^R)/\gamma_\mathrm{tot}$ with an increasing number $N=0,1,\ldots 16$ of optimization parameters. The empty squares are extrapolated values. The black solid curve shows the lower bound to the fidelity loss obtained from our analytical approach, see Eq.~(\ref{FminPAP}). The extrapolation procedure at $\lambda=0.9$ and $1$ is illustrated in panels  (b) and (c), respectively. Other parameters (setting $G_{\rm max}=1$): $\kappa^R =0.025$, $\gamma_2^R=2.5\times10^{-4}$, and $\gamma_1^\phi=\gamma_2^\phi=\kappa^\phi=0$.
}\label{PAPNumerics}
\end{figure}

\subsection{Numerical Benchmarks}

To support our variational approach, we have benchmarked it with a direct optimization of the state transfer fidelity. Similarly to Ref.~\cite{wang2017optimization}, we consider a truncated Fourier expansion of $\dot\theta(t)$:
\begin{equation}
\dot{\theta}(t)=\frac{\pi}{2t_{\mathrm{f}}}+\sum_{n=1}^N\alpha_{n}\cos\frac{n\pi t}{t_{\mathrm{f}}}\;,\label{FourierExpansion}
\end{equation}
where the constant term ($\alpha_0$) is fixed by the boundary condition $\theta(t_{\rm f})=\pi/2$.
In Fig.~\ref{Numerics} we display representative results of the numerically optimized fidelity $F_\mathrm{opt}(t_\mathrm{f})$, and compare them to our analytical framework. 

In all cases, the curves of  Fig.~\ref{Numerics} have a non-monotonic dependence on $t_{\rm f}$. The reduction at either small or large $t_\mathrm{f}$ reflects the influence of non-adiabatic leakage and direct dissipation acting on the qubit, respectively. As expected, in each panel the numerical curves increase when enlarging the space of optimization parameters (we consider $N=0,2,4$) and tend to saturate to an optimal curve, which is close to the analytical result (black solid curve). The agreement is especially good for the two lower panels, (c) and (d), where dissipation is smaller. In panels (a) and (b), the agreement is better near the maximum of the fidelity,  but visible deviations appear at smaller and larger values of $t_{\rm f}$. This is not surprising, as our formalism is based on a perturbative expansion which only retains the leading-order correction to the fidelity. Still, the agreement with numerical optimization is generally satisfactory and becomes excellent for values of the fidelity typically targeted by quantum information processing ($> 99\%$). 

We also note that, while the numerical curves of Fig.~\ref{Numerics} can at times display an oscillatory behavior (especially at small $N$), the black curves obtained from the least-action principle are smooth. In fact, as they do not constrain the pulse shape, they correspond to the limit of an infinite number of expansion coefficients in the Fourier series. Therefore, to verify more accurately the results from the least-action principle, we can perform a $N \to \infty$ extrapolation of the numerical fidelity. A comparison of these extrapolated results to the analytical treatment is presented in Fig.~\ref{PAPNumerics}, showing excellent agreement beteewn the two. In particular, one can see that the lowest solid curve of Fig.~\ref{PAPNumerics}(a), obtained from Eq.~(\ref{FminPAP}), is close to the empty squares (obtained from extrapolation).

\section{Conclusion}\label{conclusion}
In this paper, we discussed the optimization of quasi-adiabatic state-transfer protocols in dissipative systems. Through a perturbative analysis of the interplay between non-adiabatic leakage and dissipation, we expressed the fidelity loss as a classical action over the time-dependent control parameters. Therefore, our formalism leads to a mechanical analogy to the problem of quasi-adiabatic quantum state transfer and allows us to find the optimal path by applying the principle of least action. 

Furthermore, by considering the state transfer between two qubits mediated by a lossy quantum bus, we find that our treatment leads to a physically transparent interpretation in terms of a 1D particle in an external potential. This allows us to derive several analytic bounds to the fidelity, which are in excellent agreement with numerical optimization. Some direct extensions of this analysis include non-resonant $\Lambda$-systems, whose detunings are small enough compared with the system frequency-difference, and the generation of entanglement between the two qubits, similar to partial population transfer with fractional STIRAP \cite{vitanov1999creation,vitanov2017}.

The framework discussed here is immediately applicable when the eigenstates and eigenenergies of the closed-system Hamiltonian can be algebraically expressed in terms of the driving parameters. If this is not possible, the potential $V({\bf G})$ and effective mass tensor $M^{-1}({\bf G})$, defined in Eqs.~(\ref{eq:Vpotential})  and (\ref{Mtensor}), respectively, can still be obtained by relying on the numerical diagonalization of the instantaneous Hamiltonian. Therefore, we expect the method to be be generally applied to quasi-adiabatic state transfer protocols.

\begin{acknowledgments}
Y.F. acknowledges support from NSFC (Grant No. 12005011) and Yunnan Fundamental Research Projects (Grant No. 202201AU070118). Y.D.W. acknowledges support from NSFC (Grant No. 12275331) and the Penghuanwu Innovative Research Center (Grant No. 12047503). S.C. acknowledges support from the National Science Association Funds (Grant No. U2230402). Y.D.W. and S.C. acknowledge support from the Innovation Program for Quantum Science and Technology (Grant No. 2021ZD0301602).
\end{acknowledgments}

\appendix

\section{Quasi-adiabatic evolution}\label{app:Uapprox}
We discuss here the perturbative treatment of the quasi-adiabatic unitary evolution. To this end, it will be important to estimate the order of different contributions in terms of a non-adiabatic expansion parameter $\epsilon$. We denote as $G_{\max}$ the physical upper bound to the coupling constants, i.e., $G_i(t)\lesssim G_{\max}$, and estimate the energy differences as $\Delta_{m,n}\sim G_{\max}$ (if $n\neq m$). Instead, the timescale over which the couplings change (typically, between zero and $G_{\max}$) is given by $t_{\rm f}$. Finally, the $\epsilon_{m,n}$ in Eq.~(\ref{eps_mn}) can be estimated as follows:
\begin{equation}\label{epsilon}
\epsilon_{m,n} \sim  \frac{G_{\max}/t_{\rm f}}{G_{\max}^2} = \frac{1}{G_{\max}t_{\rm f}}  \ll 1.
\end{equation}
We then define $\epsilon = (G_{\max}t_{\rm f})^{-1}$, which plays the same role of the expansion parameter $v$ in the adiabatic perturbation theory of Ref.~\cite{Rigolin2008} (in our notation, $v=1/t_{\rm f}$).

We now discuss how to obtain the leading non-adiabatic terms to the wavefunction using ordinary time-dependent perturbation theory. To this end, we transform the non-adiabatic correction Hamiltonian $\hat{H}_1(t)$ in Eq.~(\ref{H0def}) to the interaction picture:
\begin{align}\label{H1I}
 \hat{H}_{1}^I(t)=&\hat{U}_0^\dagger(t)\hat{H}_{1}(t)\hat{U}_0(t)\nonumber\\
=&i\sum_n\sum_{m\neq n}\epsilon_{m,n}(t)\Delta_{m,n}(t)e^{-i\gamma_{m,n}(t)}\nonumber\\
&e^{i\phi_{m,n}(t)}|E_m(0)\rangle\langle E_n(0)|\;,
\end{align}
where $U_0(t)$ is the exact adiabatic evolution, $\gamma_{m,n}(t)=\gamma_m(t)-\gamma_n(t)$ and $\phi_{m,n}(t)=\phi_m(t)-\phi_n(t)$ denote differences of geometric and dynamical phases, respectively. The exact evolution operator is given by $\hat{U}(t)=\hat{U}_0(t)\hat{U}_{\rm I}(t)$ where, as usual, $\hat{U}_{\rm I}(t) = \mathcal{T} e^{-i\int_0^t \hat{H}_{1}^I(\tau)d\tau}$ is expanded in a power series. The first-order correction is immediately obtained as:
\begin{align}\label{U11}
\hat{U}^{(1)}_{\rm I}(t)=&\sum_n\sum_{m\neq n} \int_0^t dt_1\epsilon_{m,n}(t_1)\Delta_{m,n}(t_1) \nonumber\\
&e^{-i\gamma_{m,n}(t_1)}e^{i\phi_{m,n}(t_1)}|E_m(0)\rangle\langle E_n(0)|.
\end{align}
In the expression above, the integral is suppressed by the fast-oscillating factor $e^{i\phi_{m,n}(t_1)}$. It can be rewritten by noting that $\Delta_{m,n}(t)e^{i\phi_{m,n}(t)}=-i\frac{d}{dt}e^{i\phi_{m,n}(t)}$ and performing an integration by parts:
\begin{align}\label{U11parts}
\hat{U}^{(1)}_{\rm I}(t)=-i\sum_n\sum_{m\neq n} &\bigg(\epsilon_{m,n}(t)e^{-i\gamma_{m,n}(t)}e^{i\phi_{m,n}(t)} \nonumber\\
&-\epsilon_{m,n}(0)\bigg)|E_m(0)\rangle\langle E_n(0)| \nonumber \\
+i\sum_n\sum_{m\neq n}& \int_0^t dt_1~\frac{d}{dt_1}\left(\epsilon_{m,n}(t_1)e^{-i\gamma_{m,n}(t_1)}\right)\nonumber\\
&e^{i\phi_{m,n}(t_1)}|E_m(0)\rangle\langle E_n(0)|.
\end{align}
The second term of Eq.~(\ref{U11parts}) is much smaller than Eq.~(\ref{U11}) if the following condition is satisfied (found by comparing the integrands):
\begin{equation}\label{condition}
\left| \frac{\frac{d}{dt}\left(\epsilon_{m,n}(t)e^{-i\gamma_{m,n}(t)}\right)}{\epsilon_{m,n}(t)\Delta_{n,m}(t)}\right| \ll 1.
\end{equation}
The above Eq.~(\ref{condition}) is typically valid in the adiabatic limit since, proceeding in a way similar to Eq.~(\ref{epsilon}), the left-hand side is estimated of order $\epsilon$. Therefore, we can approximate  $\hat{U}^{(1)}$ by the first term of Eq.~(\ref{U11parts}). Conditions analogous to Eq.~(\ref{condition}) (involving higher-order time derivatives) should be imposed when extracting the leading contribution to $\hat{U}^{(k)}_{\rm I}(t)$ with $k>1$. 

While the first term of Eq.~(\ref{U11parts}) gives the correct leading contribution to the off-diagonal amplitudes, the lowest-order diagonal correction ($m=n$) is missing. To recover it, we need to consider second-order perturbation theory:
\begin{align}\label{U2}
\hat{U}^{(2)}_{\rm I}(t)=&-\int_0^t\int_0^{t_1}\hat{H}_1^I(t_1)\hat{H}_1^I(t_2)dt_2dt_1 \nonumber\\
\simeq& 
-\sum_n\sum_{m\neq n}\int_0^t dt_1\hat{H}_{1}^I(t_1)\bigg(\epsilon_{m,n}(t_1)e^{-i\gamma_{m,n}(t_1)}\nonumber\\
&e^{i\phi_{m,n}(t_1)}-\epsilon_{m,n}(0)\bigg)|E_m(0)\rangle\langle E_n(0)|.
\end{align}
where in the second step we have performed the $dt_2$ integral to leading order in $\epsilon$, in the same way discussed for $\hat{U}^{(1)}_{\rm I}(t)$. 
The crucial point to note in Eq.~(\ref{U2}) is that, for each $(n,m)$, there is a term in $\hat{H}_{1}^I(t_1)$ [see Eq.~(\ref{H1I})] which cancels the fast-oscillating factor $e^{i\phi_{m,n}(t_1)}$ appearing in Eq.~(\ref{U2}). Selecting such terms we obtain the desired diagonal correction:  
\begin{align}\label{U2approx}
\hat{U}^{(2)}_{\rm I}(t)\simeq& 
i\sum_n \sum_{m\neq n}\int_0^t dt_1~ \nonumber \\
&\Delta_{m,n}(t_1)\left|\epsilon_{m,n}(t_1)\right|^2|E_n(0)\rangle\langle E_n(0)|,
\end{align}
where we used $\epsilon_{n,m}=\epsilon_{m,n}^*$ and $\Delta_{n,m}=-\Delta_{m,n}$. Being a second-order contribution, the above expression contains a $\sim \epsilon^2$ factor in the integrand. However, due to the absence of a fast oscillating term, the long integration time (of order $t_{\rm f}$) compensates for one of such $\epsilon$ factors. The final result is of order $t_{\rm f} \Delta_{m,n} \epsilon^2 \sim \epsilon$, thus should be combined to the first term of Eq.~(\ref{U11parts}). Finally, we note that all the other contributions to $\hat{U}^{(2)}_{\rm I}(t)$, omitted in Eq.~(\ref{U2approx}), contain a fast oscillating term in the integrand, thus they are of order $\epsilon^2$ and in this work we neglect them.

Collecting previous results, we obtain the following expression for  $\hat{U}(t)=\hat{U}_0(t)\hat{U}_{\rm I}(t)$:
\begin{align}\label{Ufinal}
\hat{U}(t) = & \sum_n 
  e^{i\gamma_n(t)}e^{-i\phi_n(t)} \Bigg[ |E_n(t)\rangle\langle E_n(0)| \nonumber \\
&\times \bigg(1+ i \sum_{m\neq n}\int_0^t dt_1 \Delta_{m,n}(t_1)\left|\epsilon_{m,n}(t_1)\right|^2 + \ldots \bigg) \nonumber \\
& -i  \sum_{m\neq n}  |E_m(t)\rangle\langle E_n(0)|\nonumber \\
&\times \left(\epsilon_{m,n}(t)  -\epsilon_{m,n}(0)e^{i\gamma_{m,n}(t)}e^{-i\phi_{m,n}(t)}+\ldots \right)\Bigg].
\end{align}
Finally, we apply  $\hat{U}(t)$ to the initial condition $|E_1 (0)\rangle$ and compare the approximate evolution to the exact wavefunction in Eq.~(\ref{psit}) of the main text. To leading order in $\epsilon$, the amplitudes $c_m$ with $m\neq 1$ can be immediately found from the last line of Eq.~(\ref{Ufinal}). Furthermore, using that the leading correction to $|c_1| \simeq 1$ is of order $\epsilon^2$, we write $c_1 = |c_1|e^{i\delta\gamma_1(t)}\simeq 1 +i\delta\gamma_1(t)$ and obtain $\delta\gamma_1(t)$ from the second line of Eq.~(\ref{Ufinal}).

\begin{widetext}

\section{Derivation of the $K$ term}\label{positive_kinetic}
For the dissipative processes which satisfy Eq.~(\ref{dark_states_condtition}), we can derive Eq.~(\ref{K_final_expession}) as follows. By restricting in Eq.~(\ref{eq:fidelitylossLindblad}) the summation over $\alpha$ over the relevant terms and substituting  the general form of the unitary evolution, $|\psi (t)\rangle = e^{i\gamma_1(t)}e^{-i\phi_1(t)} \sum_m c_m(t) |E_m(t)\rangle$, we get:

\begin{align}\label{K_long_expession}
 K 
\simeq & {\sum_\alpha}^\prime \bigg[ |c_1|^2 \langle  E_1 |\hat{A}_\alpha^\dag\hat{A}_\alpha|E_1 \rangle +
\sum_{n>1}  c_n^*c_1 \langle  E_n |\hat{A}_\alpha^\dag\hat{A}_\alpha|E_1 \rangle
+\sum_{m>1}  c_1^*c_m \langle  E_1 
 |\hat{A}_\alpha^\dag\hat{A}_\alpha|E_m \rangle - |c_1|^4 \left|\langle E_1|\hat{A}_\alpha|E_1 \rangle \right|^2
\nonumber\\ &  
+ \sum_{n,m>1} c_n^*c_m \langle  E_n 
 |\hat{A}_\alpha^\dag\hat{A}_\alpha|E_m \rangle 
- |c_1|^2 \sum_{n>1}  c_n^*c_1 
\langle E_n|\hat{A}_\alpha^\dag|E_1 \rangle \langle E_1|\hat{A}_\alpha|E_1 \rangle
 - |c_1|^2 \sum_{m>1}  c_1^*c_m 
  \langle E_1|\hat{A}_\alpha^\dag|E_1 \rangle \langle E_1|\hat{A}_\alpha|E_m \rangle \nonumber\\
&- |c_1|^2 \sum_{n,m>1}  c_n^*c_m 
 \bigg( 
 \langle E_n|\hat{A}_\alpha^\dag|E_m \rangle \langle E_1|\hat{A}_\alpha|E_1 \rangle
+ \langle E_1|\hat{A}_\alpha^\dag|E_1 \rangle \langle E_n|\hat{A}_\alpha|E_m \rangle 
+ \langle E_n|\hat{A}_\alpha^\dag|E_1 \rangle \langle E_1|\hat{A}_\alpha|E_m \rangle\bigg)\bigg].
\end{align}
Above, in expanding $ |\langle\psi(t)|\hat{A}_\alpha|\psi(t)\rangle |^2$, we have omitted several terms, which are zero by virtue of Eq.~(\ref{dark_states_condtition}). We have also dropped terms which are higher order than $\epsilon^2$. Therefore, we can set $|c_1|^2 =1$ in the second and third lines of Eq.~(\ref{K_long_expession}). To further simplify Eq.~(\ref{K_long_expession}), we note that Eq.~(\ref{dark_states_condtition}) implies
$ \langle  E_{n} |\hat{A}_\alpha^\dag\hat{A}_\alpha|E_1 \rangle =  \langle  E_n |\hat{A}_\alpha^\dag|E_1\rangle\langle E_1|\hat{A}_\alpha|E_1 \rangle. $
 Thus, the $\langle  E_{n} |\hat{A}_\alpha^\dag\hat{A}_\alpha|E_1 \rangle$ term in the first line simplifies with the second term in the second line. Similar considerations hold for the $\langle  E_{1} |\hat{A}_\alpha^\dag\hat{A}_\alpha|E_m \rangle $ term, giving:
\begin{align}\label{K_long_expession2}
 K 
\simeq & {\sum_\alpha}^\prime \nonumber \bigg[ \sum_{l>1}|c_l|^2 \langle  E_1 |\hat{A}_\alpha^\dag\hat{A}_\alpha|E_1 \rangle 
+ \sum_{n,m>1} c_n^*c_m \langle  E_n 
 |\hat{A}_\alpha^\dag\hat{A}_\alpha|E_m \rangle
\nonumber\\ &
- \sum_{n,m>1}  c_n^*c_m 
 \bigg( \langle E_n|\hat{A}_\alpha^\dag|E_m \rangle \langle E_1|\hat{A}_\alpha|E_1 \rangle
 + \langle E_1|\hat{A}_\alpha^\dag|E_1 \rangle \langle E_n|\hat{A}_\alpha|E_m \rangle 
+ \langle E_n|\hat{A}_\alpha^\dag|E_1 \rangle \langle E_1|\hat{A}_\alpha|E_m \rangle\bigg)\bigg],
\end{align}
where in the first term we used $|c_1|^2(1-|c_1|^2)\simeq \sum_{l>1}|c_l|^2$. Furthermore, after inserting a completeness relation in $\langle  E_n 
 |\hat{A}_\alpha^\dag\hat{A}_\alpha|E_m \rangle$ (first line), we cancel the third term in the round parenthesis (second line):
 \begin{align}\label{K_long_expession3}
 K \simeq & {\sum_\alpha}^\prime \bigg[ \sum_{l>1} |c_l|^2 \langle  E_1 |\hat{A}_\alpha^\dag |  E_1 \rangle\langle  E_1 |\hat{A}_\alpha|E_1 \rangle
 +   \sum_{l,n,m>1} c_n^*c_m \langle  E_n 
 |\hat{A}_\alpha^\dag |  E_l \rangle\langle  E_l |\hat{A}_\alpha|E_m \rangle\nonumber\\
&- \sum_{n,m>1}  c_n^*c_m \bigg( \langle E_n|\hat{A}_\alpha^\dag|E_m \rangle \langle E_1|\hat{A}_\alpha|E_1 \rangle
+ \langle E_1|\hat{A}_\alpha^\dag|E_1 \rangle \langle E_n|\hat{A}_\alpha|E_m \rangle \bigg )\bigg] .
\end{align}
Finally, by relabeling the summation indices, the last expression can be written as:
\begin{align}
\label{K_long_expession4}
 K 
\simeq & {\sum_\alpha}^\prime \sum_{l>1} \bigg[  c_l^* \langle  E_1 |\hat{A}_\alpha^\dag |  E_1 \rangle c_l \langle  E_1 |\hat{A}_\alpha|E_1 \rangle 
+  \bigg( \sum_{n>1} c_n^*\langle  E_n 
 |\hat{A}_\alpha^\dag |  E_l \rangle \bigg)\bigg( \sum_{m>1}c_m \langle  E_l |\hat{A}_\alpha|E_m \rangle \bigg)\nonumber \\
&-  \bigg(\sum_{n>1}  c_n^* 
 \langle E_n|\hat{A}_\alpha^\dag|E_l \rangle \bigg) c_l \langle E_1|\hat{A}_\alpha|E_1 \rangle 
-  
 c_l^* \langle E_1|\hat{A}_\alpha^\dag|E_1 \rangle \bigg( \sum_{m>1} c_m \langle E_l|\hat{A}_\alpha|E_m \rangle\bigg) \bigg] ,
\end{align}
  which coincides with Eq.~(\ref{K_final_expession}) of the main text.

  \end{widetext}

  \section{Detailed considerations about the boundary condition}\label{app:boundaryc}
  
In analysing the optimal fidelity, we have set the boundary condition in Eq.~(\ref{dGdt}). However, a general problem is that this requirement may be inconsistent with the dynamcs of the associated classical system. For example, for the case of Sec.~\ref{Model and perturbation}, Eq.~(\ref{dGdt}) implies that the mixing angle $\theta(t)$ has vanishing time derivatives at the boundaries $\theta=0$ and $\pi/2$. However, the optimization procedures in Sec.~\ref{optimization} show that the value of $\dot\theta$ depends on the transfer time through the energy $\mathcal{E}$, see Eq.~(\ref{dtheta}), and in general $\dot{\theta}\neq 0$ when $\theta=0,\pi/2$. This is especially true when one performs an optimization with respect to $\mathcal{E}$ (or, equivalently, $t_{\rm f}$). 

For example, consider the PAP case (where $G(t)=G_0$ is a constant) with $\gamma^R_1=\gamma^R_2$ and $\gamma^\phi_1=\gamma^\phi_2=0$. The explicit solution to the Euler-Lagrange equation is
\begin{equation}\label{PAPspec}
\theta(t)=\frac{\pi t}{2t_{\mathrm{f}}}\;,
\end{equation}
which does not satisfy $\dot{\theta}(0)=\dot{\theta}(t_{\rm f})=0$. Formally, however, one can easily find a modification of Eq.~(\ref{PAPspec}) which satisfies $\dot\theta(0)=\dot\theta(t_f)=0$ and approaches with arbitrary precision the optimal fidelity of the Lagrangian formalism. For example:
\begin{equation}
\dot{\theta}(t)=\begin{cases}
\frac{\pi t}{2t_{\mathrm{f}}\Delta t} & ,\text{ for }0\leq t<\Delta t\\
\frac{\pi}{2t_{\mathrm{f}}} & ,\text{ for }\Delta t\leq t<t_{\mathrm{f}}-\Delta t\\
\frac{\pi}{2t_{\mathrm{f}}}-\frac{\pi t}{2t_{\mathrm{f}}\Delta t} & ,\text{ for }t_{\mathrm{f}}-\Delta t\leq t\leq t_{\mathrm{f}}
\end{cases}\label{thetasmooth}
\end{equation}
where $\Delta t$ is a sufficiently small time interval.  In fact, a form analogous to Eq.~(\ref{thetasmooth}) was found in Ref.~\cite{wang2017optimization}, where additional boundary terms in the perturbative expansion were retained.

In general we expect that, accounting for the full dynamical evolution, the optimal $\theta(t)$ will differ from Eq.~(\ref{PAPspec}) in the detailed dependence around $t=0,t_{\rm f}$. This might introduce some difficulties from the point of view of the non-adiabatic pertubative expansion. Notice in particular, that the modified $\dot{\theta}(t)$ in Eq.~(\ref{thetasmooth}) will violate the adiabatic condition, as $\ddot{\theta}\propto 1/\Delta t$ approaches infinity when $\Delta t \to 0$. Nevertheless, the influence on the fidelity should remain small. This is because the difference $||\rho(0)-\rho(\Delta t)||_\infty$ will be restricted by the scale of $\Delta t$, which is expected to be small compared to the total evolution time $t_{\rm f}$. The comparisons to explicit numerical optimization, indicate that these corrections indeed have a small influence.


\bibliographystyle{apsrev4-2}

\begin{thebibliography}{55}%
\makeatletter
\providecommand \@ifxundefined [1]{%
 \@ifx{#1\undefined}
}%
\providecommand \@ifnum [1]{%
 \ifnum #1\expandafter \@firstoftwo
 \else \expandafter \@secondoftwo
 \fi
}%
\providecommand \@ifx [1]{%
 \ifx #1\expandafter \@firstoftwo
 \else \expandafter \@secondoftwo
 \fi
}%
\providecommand \natexlab [1]{#1}%
\providecommand \enquote  [1]{``#1''}%
\providecommand \bibnamefont  [1]{#1}%
\providecommand \bibfnamefont [1]{#1}%
\providecommand \citenamefont [1]{#1}%
\providecommand \href@noop [0]{\@secondoftwo}%
\providecommand \href [0]{\begingroup \@sanitize@url \@href}%
\providecommand \@href[1]{\@@startlink{#1}\@@href}%
\providecommand \@@href[1]{\endgroup#1\@@endlink}%
\providecommand \@sanitize@url [0]{\catcode `\\12\catcode `\$12\catcode `\&12\catcode `\#12\catcode `\^12\catcode `\_12\catcode `\%12\relax}%
\providecommand \@@startlink[1]{}%
\providecommand \@@endlink[0]{}%
\providecommand \url  [0]{\begingroup\@sanitize@url \@url }%
\providecommand \@url [1]{\endgroup\@href {#1}{\urlprefix }}%
\providecommand \urlprefix  [0]{URL }%
\providecommand \Eprint [0]{\href }%
\providecommand \doibase [0]{https://doi.org/}%
\providecommand \selectlanguage [0]{\@gobble}%
\providecommand \bibinfo  [0]{\@secondoftwo}%
\providecommand \bibfield  [0]{\@secondoftwo}%
\providecommand \translation [1]{[#1]}%
\providecommand \BibitemOpen [0]{}%
\providecommand \bibitemStop [0]{}%
\providecommand \bibitemNoStop [0]{.\EOS\space}%
\providecommand \EOS [0]{\spacefactor3000\relax}%
\providecommand \BibitemShut  [1]{\csname bibitem#1\endcsname}%
\let\auto@bib@innerbib\@empty
\bibitem [{\citenamefont {Peirce}\ \emph {et~al.}(1988)\citenamefont {Peirce}, \citenamefont {Dahleh},\ and\ \citenamefont {Rabitz}}]{peirce1988}%
  \BibitemOpen
  \bibfield  {author} {\bibinfo {author} {\bibfnamefont {A.~P.}\ \bibnamefont {Peirce}}, \bibinfo {author} {\bibfnamefont {M.~A.}\ \bibnamefont {Dahleh}},\ and\ \bibinfo {author} {\bibfnamefont {H.}~\bibnamefont {Rabitz}},\ }\bibfield  {title} {\bibinfo {title} {Optimal control of quantum-mechanical systems: Existence, numerical approximation, and applications},\ }\href {https://doi.org/10.1103/PhysRevA.3u7.4950} {\bibfield  {journal} {\bibinfo  {journal} {Phys. Rev. A}\ }\textbf {\bibinfo {volume} {37}},\ \bibinfo {pages} {4950--4964} (\bibinfo {year} {1988})}\BibitemShut {NoStop}%
\bibitem [{\citenamefont {Ohtsuki}\ \emph {et~al.}(1999)\citenamefont {Ohtsuki}, \citenamefont {Zhu},\ and\ \citenamefont {Rabitz}}]{ohtsuki1999}%
  \BibitemOpen
  \bibfield  {author} {\bibinfo {author} {\bibfnamefont {Y.}~\bibnamefont {Ohtsuki}}, \bibinfo {author} {\bibfnamefont {W.}~\bibnamefont {Zhu}},\ and\ \bibinfo {author} {\bibfnamefont {H.}~\bibnamefont {Rabitz}},\ }\bibfield  {title} {\bibinfo {title} {Monotonically convergent algorithm for quantum optimal control with dissipation},\ }\href {https://doi.org/10.1063/1.478036} {\bibfield  {journal} {\bibinfo  {journal} {The Journal of Chemical Physics}\ }\textbf {\bibinfo {volume} {110}},\ \bibinfo {pages} {9825--9832} (\bibinfo {year} {1999})}\BibitemShut {NoStop}%
\bibitem [{\citenamefont {Werschnik}\ and\ \citenamefont {Gross}(2007)}]{Werschnik2007QuantumOC}%
  \BibitemOpen
  \bibfield  {author} {\bibinfo {author} {\bibfnamefont {J.}~\bibnamefont {Werschnik}}\ and\ \bibinfo {author} {\bibfnamefont {E.~K.~U.}\ \bibnamefont {Gross}},\ }\bibfield  {title} {\bibinfo {title} {Quantum optimal control theory},\ }\href {https://api.semanticscholar.org/CorpusID:1595608} {\bibfield  {journal} {\bibinfo  {journal} {Journal of Physics B: Atomic, Molecular and Optical Physics}\ }\textbf {\bibinfo {volume} {40}},\ \bibinfo {pages} {R175 -- R211} (\bibinfo {year} {2007})}\BibitemShut {NoStop}%
\bibitem [{\citenamefont {Vitanov}\ \emph {et~al.}(2001)\citenamefont {Vitanov}, \citenamefont {Halfmann}, \citenamefont {Shore},\ and\ \citenamefont {Bergmann}}]{adiabatic_protocols}%
  \BibitemOpen
  \bibfield  {author} {\bibinfo {author} {\bibfnamefont {N.~V.}\ \bibnamefont {Vitanov}}, \bibinfo {author} {\bibfnamefont {T.}~\bibnamefont {Halfmann}}, \bibinfo {author} {\bibfnamefont {B.~W.}\ \bibnamefont {Shore}},\ and\ \bibinfo {author} {\bibfnamefont {K.}~\bibnamefont {Bergmann}},\ }\bibfield  {title} {\bibinfo {title} {Laser-induced population transfer by adiabatic passage techniques},\ }\href {https://doi.org/10.1146/annurev.physchem.52.1.763} {\bibfield  {journal} {\bibinfo  {journal} {Annual Review of Physical Chemistry}\ }\textbf {\bibinfo {volume} {52}},\ \bibinfo {pages} {763--809} (\bibinfo {year} {2001})}\BibitemShut {NoStop}%
\bibitem [{\citenamefont {Emmanouilidou}\ \emph {et~al.}(2000)\citenamefont {Emmanouilidou}, \citenamefont {Zhao}, \citenamefont {Ao},\ and\ \citenamefont {Niu}}]{Emmanouilidou2000}%
  \BibitemOpen
  \bibfield  {author} {\bibinfo {author} {\bibfnamefont {A.}~\bibnamefont {Emmanouilidou}}, \bibinfo {author} {\bibfnamefont {X.-G.}\ \bibnamefont {Zhao}}, \bibinfo {author} {\bibfnamefont {P.}~\bibnamefont {Ao}},\ and\ \bibinfo {author} {\bibfnamefont {Q.}~\bibnamefont {Niu}},\ }\bibfield  {title} {\bibinfo {title} {Steering an eigenstate to a destination},\ }\href {https://doi.org/10.1103/PhysRevLett.85.1626} {\bibfield  {journal} {\bibinfo  {journal} {Phys. Rev. Lett.}\ }\textbf {\bibinfo {volume} {85}},\ \bibinfo {pages} {1626--1629} (\bibinfo {year} {2000})}\BibitemShut {NoStop}%
\bibitem [{\citenamefont {Torrontegui}\ \emph {et~al.}(2013)\citenamefont {Torrontegui}, \citenamefont {Ibáñez}, \citenamefont {Martínez-Garaot}, \citenamefont {Modugno}, \citenamefont {del Campo}, \citenamefont {Guéry-Odelin}, \citenamefont {Ruschhaupt}, \citenamefont {Chen},\ and\ \citenamefont {Muga}}]{Torrontegui2013}%
  \BibitemOpen
  \bibfield  {author} {\bibinfo {author} {\bibfnamefont {E.}~\bibnamefont {Torrontegui}}, \bibinfo {author} {\bibfnamefont {S.}~\bibnamefont {Ibáñez}}, \bibinfo {author} {\bibfnamefont {S.}~\bibnamefont {Martínez-Garaot}}, \bibinfo {author} {\bibfnamefont {M.}~\bibnamefont {Modugno}}, \bibinfo {author} {\bibfnamefont {A.}~\bibnamefont {del Campo}}, \bibinfo {author} {\bibfnamefont {D.}~\bibnamefont {Guéry-Odelin}}, \bibinfo {author} {\bibfnamefont {A.}~\bibnamefont {Ruschhaupt}}, \bibinfo {author} {\bibfnamefont {X.}~\bibnamefont {Chen}},\ and\ \bibinfo {author} {\bibfnamefont {J.~G.}\ \bibnamefont {Muga}},\ }\bibinfo {title} {Shortcuts to adiabaticity},\ in\ \href {https://doi.org/10.1016/b978-0-12-408090-4.00002-5} {\emph {\bibinfo {booktitle} {Advances in Atomic, Molecular, and Optical Physics}}}\ (\bibinfo  {publisher} {Elsevier},\ \bibinfo {year} {2013})\ p.\ \bibinfo {pages} {117–169}\BibitemShut {NoStop}%
\bibitem [{\citenamefont {Gu\'ery-Odelin}\ \emph {et~al.}(2019)\citenamefont {Gu\'ery-Odelin}, \citenamefont {Ruschhaupt}, \citenamefont {Kiely}, \citenamefont {Torrontegui}, \citenamefont {Mart\'{\i}nez-Garaot},\ and\ \citenamefont {Muga}}]{guery2019}%
  \BibitemOpen
  \bibfield  {author} {\bibinfo {author} {\bibfnamefont {D.}~\bibnamefont {Gu\'ery-Odelin}}, \bibinfo {author} {\bibfnamefont {A.}~\bibnamefont {Ruschhaupt}}, \bibinfo {author} {\bibfnamefont {A.}~\bibnamefont {Kiely}}, \bibinfo {author} {\bibfnamefont {E.}~\bibnamefont {Torrontegui}}, \bibinfo {author} {\bibfnamefont {S.}~\bibnamefont {Mart\'{\i}nez-Garaot}},\ and\ \bibinfo {author} {\bibfnamefont {J.~G.}\ \bibnamefont {Muga}},\ }\bibfield  {title} {\bibinfo {title} {Shortcuts to adiabaticity: Concepts, methods, and applications},\ }\href {https://doi.org/10.1103/RevModPhys.91.045001} {\bibfield  {journal} {\bibinfo  {journal} {Rev. Mod. Phys.}\ }\textbf {\bibinfo {volume} {91}},\ \bibinfo {pages} {045001} (\bibinfo {year} {2019})}\BibitemShut {NoStop}%
\bibitem [{\citenamefont {{Unanyan}}\ \emph {et~al.}(1997)\citenamefont {{Unanyan}}, \citenamefont {{Yatsenko}}, \citenamefont {{Bergmann}},\ and\ \citenamefont {{Shore}}}]{shore1997}%
  \BibitemOpen
  \bibfield  {author} {\bibinfo {author} {\bibfnamefont {R.~G.}\ \bibnamefont {{Unanyan}}}, \bibinfo {author} {\bibfnamefont {L.~P.}\ \bibnamefont {{Yatsenko}}}, \bibinfo {author} {\bibfnamefont {K.}~\bibnamefont {{Bergmann}}},\ and\ \bibinfo {author} {\bibfnamefont {B.~W.}\ \bibnamefont {{Shore}}},\ }\bibfield  {title} {\bibinfo {title} {{Laser-induced adiabatic atomic reorientation with control of diabatic losses}},\ }\href {https://doi.org/10.1016/S0030-4018(97)00099-0} {\bibfield  {journal} {\bibinfo  {journal} {Optics Communications}\ }\textbf {\bibinfo {volume} {139}},\ \bibinfo {pages} {48--54} (\bibinfo {year} {1997})}\BibitemShut {NoStop}%
\bibitem [{\citenamefont {Fleischhauer}\ \emph {et~al.}(1999)\citenamefont {Fleischhauer}, \citenamefont {Unanyan}, \citenamefont {Shore},\ and\ \citenamefont {Bergmann}}]{bergmann1999}%
  \BibitemOpen
  \bibfield  {author} {\bibinfo {author} {\bibfnamefont {M.}~\bibnamefont {Fleischhauer}}, \bibinfo {author} {\bibfnamefont {R.}~\bibnamefont {Unanyan}}, \bibinfo {author} {\bibfnamefont {B.~W.}\ \bibnamefont {Shore}},\ and\ \bibinfo {author} {\bibfnamefont {K.}~\bibnamefont {Bergmann}},\ }\bibfield  {title} {\bibinfo {title} {Coherent population transfer beyond the adiabatic limit: Generalized matched pulses and higher-order trapping states},\ }\href {https://doi.org/10.1103/PhysRevA.59.3751} {\bibfield  {journal} {\bibinfo  {journal} {Phys. Rev. A}\ }\textbf {\bibinfo {volume} {59}},\ \bibinfo {pages} {3751--3760} (\bibinfo {year} {1999})}\BibitemShut {NoStop}%
\bibitem [{\citenamefont {Demirplak}\ and\ \citenamefont {Rice}(2008)}]{demirplak_consistency_2008}%
  \BibitemOpen
  \bibfield  {author} {\bibinfo {author} {\bibfnamefont {M.}~\bibnamefont {Demirplak}}\ and\ \bibinfo {author} {\bibfnamefont {S.~A.}\ \bibnamefont {Rice}},\ }\bibfield  {title} {\bibinfo {title} {On the consistency, extremal, and global properties of counterdiabatic fields},\ }\href {https://doi.org/10.1063/1.2992152} {\bibfield  {journal} {\bibinfo  {journal} {The Journal of Chemical Physics}\ }\textbf {\bibinfo {volume} {129}},\ \bibinfo {pages} {154111} (\bibinfo {year} {2008})}\BibitemShut {NoStop}%
\bibitem [{\citenamefont {Berry}(2009)}]{Berry_2009}%
  \BibitemOpen
  \bibfield  {author} {\bibinfo {author} {\bibfnamefont {M.~V.}\ \bibnamefont {Berry}},\ }\bibfield  {title} {\bibinfo {title} {Transitionless quantum driving},\ }\href {https://doi.org/10.1088/1751-8113/42/36/365303} {\bibfield  {journal} {\bibinfo  {journal} {Journal of Physics A: Mathematical and Theoretical}\ }\textbf {\bibinfo {volume} {42}},\ \bibinfo {pages} {365303} (\bibinfo {year} {2009})}\BibitemShut {NoStop}%
\bibitem [{\citenamefont {del Campo}(2013)}]{delcampo2013}%
  \BibitemOpen
  \bibfield  {author} {\bibinfo {author} {\bibfnamefont {A.}~\bibnamefont {del Campo}},\ }\bibfield  {title} {\bibinfo {title} {Shortcuts to adiabaticity by counterdiabatic driving},\ }\href {https://doi.org/10.1103/PhysRevLett.111.100502} {\bibfield  {journal} {\bibinfo  {journal} {Phys. Rev. Lett.}\ }\textbf {\bibinfo {volume} {111}},\ \bibinfo {pages} {100502} (\bibinfo {year} {2013})}\BibitemShut {NoStop}%
\bibitem [{\citenamefont {Petiziol}\ \emph {et~al.}(2020)\citenamefont {Petiziol}, \citenamefont {Arimondo}, \citenamefont {Giannelli}, \citenamefont {Mintert},\ and\ \citenamefont {Wimberger}}]{petiziol_optimized_2020}%
  \BibitemOpen
  \bibfield  {author} {\bibinfo {author} {\bibfnamefont {F.}~\bibnamefont {Petiziol}}, \bibinfo {author} {\bibfnamefont {E.}~\bibnamefont {Arimondo}}, \bibinfo {author} {\bibfnamefont {L.}~\bibnamefont {Giannelli}}, \bibinfo {author} {\bibfnamefont {F.}~\bibnamefont {Mintert}},\ and\ \bibinfo {author} {\bibfnamefont {S.}~\bibnamefont {Wimberger}},\ }\bibfield  {title} {\bibinfo {title} {Optimized three-level quantum transfers based on frequency-modulated optical excitations},\ }\href {https://doi.org/10.1038/s41598-020-59046-8} {\bibfield  {journal} {\bibinfo  {journal} {Scientific Reports}\ }\textbf {\bibinfo {volume} {10}},\ \bibinfo {pages} {2185} (\bibinfo {year} {2020})}\BibitemShut {NoStop}%
\bibitem [{\citenamefont {Verdeny}\ \emph {et~al.}(2014)\citenamefont {Verdeny}, \citenamefont {Rudnicki}, \citenamefont {M\"uller},\ and\ \citenamefont {Mintert}}]{verdeny2014}%
  \BibitemOpen
  \bibfield  {author} {\bibinfo {author} {\bibfnamefont {A.}~\bibnamefont {Verdeny}}, \bibinfo {author} {\bibfnamefont {L.}~\bibnamefont {Rudnicki}}, \bibinfo {author} {\bibfnamefont {C.~A.}\ \bibnamefont {M\"uller}},\ and\ \bibinfo {author} {\bibfnamefont {F.}~\bibnamefont {Mintert}},\ }\bibfield  {title} {\bibinfo {title} {Optimal control of effective hamiltonians},\ }\href {https://doi.org/10.1103/PhysRevLett.113.010501} {\bibfield  {journal} {\bibinfo  {journal} {Phys. Rev. Lett.}\ }\textbf {\bibinfo {volume} {113}},\ \bibinfo {pages} {010501} (\bibinfo {year} {2014})}\BibitemShut {NoStop}%
\bibitem [{\citenamefont {Baksic}\ \emph {et~al.}(2016)\citenamefont {Baksic}, \citenamefont {Ribeiro},\ and\ \citenamefont {Clerk}}]{baksic2016speeding}%
  \BibitemOpen
  \bibfield  {author} {\bibinfo {author} {\bibfnamefont {A.}~\bibnamefont {Baksic}}, \bibinfo {author} {\bibfnamefont {H.}~\bibnamefont {Ribeiro}},\ and\ \bibinfo {author} {\bibfnamefont {A.~A.}\ \bibnamefont {Clerk}},\ }\bibfield  {title} {\bibinfo {title} {Speeding up adiabatic quantum state transfer by using dressed states},\ }\href {https://doi.org/10.1103/PhysRevLett.116.230503} {\bibfield  {journal} {\bibinfo  {journal} {Phys. Rev. Lett.}\ }\textbf {\bibinfo {volume} {116}},\ \bibinfo {pages} {230503} (\bibinfo {year} {2016})}\BibitemShut {NoStop}%
\bibitem [{\citenamefont {Ribeiro}\ \emph {et~al.}(2017)\citenamefont {Ribeiro}, \citenamefont {Baksic},\ and\ \citenamefont {Clerk}}]{ribeiro2017systematic}%
  \BibitemOpen
  \bibfield  {author} {\bibinfo {author} {\bibfnamefont {H.}~\bibnamefont {Ribeiro}}, \bibinfo {author} {\bibfnamefont {A.}~\bibnamefont {Baksic}},\ and\ \bibinfo {author} {\bibfnamefont {A.~A.}\ \bibnamefont {Clerk}},\ }\bibfield  {title} {\bibinfo {title} {Systematic magnus-based approach for suppressing leakage and nonadiabatic errors in quantum dynamics},\ }\href {https://doi.org/10.1103/PhysRevX.7.011021} {\bibfield  {journal} {\bibinfo  {journal} {Phys. Rev. X}\ }\textbf {\bibinfo {volume} {7}},\ \bibinfo {pages} {011021} (\bibinfo {year} {2017})}\BibitemShut {NoStop}%
\bibitem [{\citenamefont {Evangelakos}\ \emph {et~al.}(2023)\citenamefont {Evangelakos}, \citenamefont {Paspalakis},\ and\ \citenamefont {Stefanatos}}]{Evangelakos2023}%
  \BibitemOpen
  \bibfield  {author} {\bibinfo {author} {\bibfnamefont {V.}~\bibnamefont {Evangelakos}}, \bibinfo {author} {\bibfnamefont {E.}~\bibnamefont {Paspalakis}},\ and\ \bibinfo {author} {\bibfnamefont {D.}~\bibnamefont {Stefanatos}},\ }\bibfield  {title} {\bibinfo {title} {Optimal stirap shortcuts using the spin-to-spring mapping},\ }\href {https://doi.org/10.1103/PhysRevA.107.052606} {\bibfield  {journal} {\bibinfo  {journal} {Phys. Rev. A}\ }\textbf {\bibinfo {volume} {107}},\ \bibinfo {pages} {052606} (\bibinfo {year} {2023})}\BibitemShut {NoStop}%
\bibitem [{\citenamefont {Greentree}\ \emph {et~al.}(2004)\citenamefont {Greentree}, \citenamefont {Cole}, \citenamefont {Hamilton},\ and\ \citenamefont {Hollenberg}}]{greentree2004coherent}%
  \BibitemOpen
  \bibfield  {author} {\bibinfo {author} {\bibfnamefont {A.~D.}\ \bibnamefont {Greentree}}, \bibinfo {author} {\bibfnamefont {J.~H.}\ \bibnamefont {Cole}}, \bibinfo {author} {\bibfnamefont {A.~R.}\ \bibnamefont {Hamilton}},\ and\ \bibinfo {author} {\bibfnamefont {L.~C.~L.}\ \bibnamefont {Hollenberg}},\ }\bibfield  {title} {\bibinfo {title} {Coherent electronic transfer in quantum dot systems using adiabatic passage},\ }\href {https://doi.org/10.1103/PhysRevB.70.235317} {\bibfield  {journal} {\bibinfo  {journal} {Phys. Rev. B}\ }\textbf {\bibinfo {volume} {70}},\ \bibinfo {pages} {235317} (\bibinfo {year} {2004})}\BibitemShut {NoStop}%
\bibitem [{\citenamefont {Ivanov}\ \emph {et~al.}(2005)\citenamefont {Ivanov}, \citenamefont {Vitanov},\ and\ \citenamefont {Bergmann}}]{ivanov2005spontaneous}%
  \BibitemOpen
  \bibfield  {author} {\bibinfo {author} {\bibfnamefont {P.~A.}\ \bibnamefont {Ivanov}}, \bibinfo {author} {\bibfnamefont {N.~V.}\ \bibnamefont {Vitanov}},\ and\ \bibinfo {author} {\bibfnamefont {K.}~\bibnamefont {Bergmann}},\ }\bibfield  {title} {\bibinfo {title} {Spontaneous emission in stimulated raman adiabatic passage},\ }\href {https://doi.org/10.1103/PhysRevA.72.053412} {\bibfield  {journal} {\bibinfo  {journal} {Phys. Rev. A}\ }\textbf {\bibinfo {volume} {72}},\ \bibinfo {pages} {053412} (\bibinfo {year} {2005})}\BibitemShut {NoStop}%
\bibitem [{\citenamefont {Goto}\ and\ \citenamefont {Ichimura}(2008)}]{goto2008upper}%
  \BibitemOpen
  \bibfield  {author} {\bibinfo {author} {\bibfnamefont {H.}~\bibnamefont {Goto}}\ and\ \bibinfo {author} {\bibfnamefont {K.}~\bibnamefont {Ichimura}},\ }\bibfield  {title} {\bibinfo {title} {Upper bound for the success probability of cavity-mediated adiabatic transfer in the presence of dissipation},\ }\href {https://doi.org/10.1103/PhysRevA.77.013816} {\bibfield  {journal} {\bibinfo  {journal} {Phys. Rev. A}\ }\textbf {\bibinfo {volume} {77}},\ \bibinfo {pages} {013816} (\bibinfo {year} {2008})}\BibitemShut {NoStop}%
\bibitem [{\citenamefont {Scala}\ \emph {et~al.}(2010)\citenamefont {Scala}, \citenamefont {Militello}, \citenamefont {Messina},\ and\ \citenamefont {Vitanov}}]{scala2010stimulated}%
  \BibitemOpen
  \bibfield  {author} {\bibinfo {author} {\bibfnamefont {M.}~\bibnamefont {Scala}}, \bibinfo {author} {\bibfnamefont {B.}~\bibnamefont {Militello}}, \bibinfo {author} {\bibfnamefont {A.}~\bibnamefont {Messina}},\ and\ \bibinfo {author} {\bibfnamefont {N.~V.}\ \bibnamefont {Vitanov}},\ }\bibfield  {title} {\bibinfo {title} {Stimulated raman adiabatic passage in an open quantum system: Master equation approach},\ }\href {https://doi.org/10.1103/PhysRevA.81.053847} {\bibfield  {journal} {\bibinfo  {journal} {Phys. Rev. A}\ }\textbf {\bibinfo {volume} {81}},\ \bibinfo {pages} {053847} (\bibinfo {year} {2010})}\BibitemShut {NoStop}%
\bibitem [{\citenamefont {Vogt}\ \emph {et~al.}(2012)\citenamefont {Vogt}, \citenamefont {Cole}, \citenamefont {Marthaler},\ and\ \citenamefont {Sch\"on}}]{vogt2012influence}%
  \BibitemOpen
  \bibfield  {author} {\bibinfo {author} {\bibfnamefont {N.}~\bibnamefont {Vogt}}, \bibinfo {author} {\bibfnamefont {J.~H.}\ \bibnamefont {Cole}}, \bibinfo {author} {\bibfnamefont {M.}~\bibnamefont {Marthaler}},\ and\ \bibinfo {author} {\bibfnamefont {G.}~\bibnamefont {Sch\"on}},\ }\bibfield  {title} {\bibinfo {title} {Influence of two-level fluctuators on adiabatic passage techniques},\ }\href {https://doi.org/10.1103/PhysRevB.85.174515} {\bibfield  {journal} {\bibinfo  {journal} {Phys. Rev. B}\ }\textbf {\bibinfo {volume} {85}},\ \bibinfo {pages} {174515} (\bibinfo {year} {2012})}\BibitemShut {NoStop}%
\bibitem [{\citenamefont {Yuan}\ \emph {et~al.}(2012)\citenamefont {Yuan}, \citenamefont {Koch}, \citenamefont {Salamon},\ and\ \citenamefont {Tannor}}]{yuan2012controllability}%
  \BibitemOpen
  \bibfield  {author} {\bibinfo {author} {\bibfnamefont {H.}~\bibnamefont {Yuan}}, \bibinfo {author} {\bibfnamefont {C.~P.}\ \bibnamefont {Koch}}, \bibinfo {author} {\bibfnamefont {P.}~\bibnamefont {Salamon}},\ and\ \bibinfo {author} {\bibfnamefont {D.~J.}\ \bibnamefont {Tannor}},\ }\bibfield  {title} {\bibinfo {title} {Controllability on relaxation-free subspaces: On the relationship between adiabatic population transfer and optimal control},\ }\href {https://doi.org/10.1103/PhysRevA.85.033417} {\bibfield  {journal} {\bibinfo  {journal} {Phys. Rev. A}\ }\textbf {\bibinfo {volume} {85}},\ \bibinfo {pages} {033417} (\bibinfo {year} {2012})}\BibitemShut {NoStop}%
\bibitem [{\citenamefont {Hou}\ \emph {et~al.}(2013)\citenamefont {Hou}, \citenamefont {Yang}, \citenamefont {Feng},\ and\ \citenamefont {Chen}}]{hou2013quantum}%
  \BibitemOpen
  \bibfield  {author} {\bibinfo {author} {\bibfnamefont {Q.~Z.}\ \bibnamefont {Hou}}, \bibinfo {author} {\bibfnamefont {W.~L.}\ \bibnamefont {Yang}}, \bibinfo {author} {\bibfnamefont {M.}~\bibnamefont {Feng}},\ and\ \bibinfo {author} {\bibfnamefont {C.-Y.}\ \bibnamefont {Chen}},\ }\bibfield  {title} {\bibinfo {title} {Quantum state transfer using stimulated raman adiabatic passage under a dissipative environment},\ }\href {https://doi.org/10.1103/PhysRevA.88.013807} {\bibfield  {journal} {\bibinfo  {journal} {Phys. Rev. A}\ }\textbf {\bibinfo {volume} {88}},\ \bibinfo {pages} {013807} (\bibinfo {year} {2013})}\BibitemShut {NoStop}%
\bibitem [{\citenamefont {Dupont-Nivet}\ \emph {et~al.}(2015)\citenamefont {Dupont-Nivet}, \citenamefont {Casiulis}, \citenamefont {Laudat}, \citenamefont {Westbrook},\ and\ \citenamefont {Schwartz}}]{dupont2015microwave}%
  \BibitemOpen
  \bibfield  {author} {\bibinfo {author} {\bibfnamefont {M.}~\bibnamefont {Dupont-Nivet}}, \bibinfo {author} {\bibfnamefont {M.}~\bibnamefont {Casiulis}}, \bibinfo {author} {\bibfnamefont {T.}~\bibnamefont {Laudat}}, \bibinfo {author} {\bibfnamefont {C.~I.}\ \bibnamefont {Westbrook}},\ and\ \bibinfo {author} {\bibfnamefont {S.}~\bibnamefont {Schwartz}},\ }\bibfield  {title} {\bibinfo {title} {Microwave-stimulated raman adiabatic passage in a bose-einstein condensate on an atom chip},\ }\href {https://doi.org/10.1103/PhysRevA.91.053420} {\bibfield  {journal} {\bibinfo  {journal} {Phys. Rev. A}\ }\textbf {\bibinfo {volume} {91}},\ \bibinfo {pages} {053420} (\bibinfo {year} {2015})}\BibitemShut {NoStop}%
\bibitem [{\citenamefont {Wang}\ \emph {et~al.}(2016)\citenamefont {Wang}, \citenamefont {Zhang}, \citenamefont {Yan},\ and\ \citenamefont {Chesi}}]{wang2017optimization}%
  \BibitemOpen
  \bibfield  {author} {\bibinfo {author} {\bibfnamefont {Y.-D.}\ \bibnamefont {Wang}}, \bibinfo {author} {\bibfnamefont {R.}~\bibnamefont {Zhang}}, \bibinfo {author} {\bibfnamefont {X.-B.}\ \bibnamefont {Yan}},\ and\ \bibinfo {author} {\bibfnamefont {S.}~\bibnamefont {Chesi}},\ }\bibfield  {title} {\bibinfo {title} {Optimization of stirap-based state transfer under dissipation},\ }\href {https://api.semanticscholar.org/CorpusID:118515937} {\bibfield  {journal} {\bibinfo  {journal} {New Journal of Physics}\ }\textbf {\bibinfo {volume} {19}} (\bibinfo {year} {2016})}\BibitemShut {NoStop}%
\bibitem [{\citenamefont {Stefanatos}\ and\ \citenamefont {Paspalakis}(2021)}]{stefanatos2021}%
  \BibitemOpen
  \bibfield  {author} {\bibinfo {author} {\bibfnamefont {D.}~\bibnamefont {Stefanatos}}\ and\ \bibinfo {author} {\bibfnamefont {E.}~\bibnamefont {Paspalakis}},\ }\bibfield  {title} {\bibinfo {title} {Optimal shape of {STIRAP} pulses for large dissipation at the intermediate level},\ }\href {https://doi.org/10.1007/s11128-021-03352-1} {\bibfield  {journal} {\bibinfo  {journal} {Quantum Information Processing}\ }\textbf {\bibinfo {volume} {20}},\ \bibinfo {pages} {391} (\bibinfo {year} {2021})}\BibitemShut {NoStop}%
\bibitem [{\citenamefont {Gaubatz}\ \emph {et~al.}(1988)\citenamefont {Gaubatz}, \citenamefont {Rudecki}, \citenamefont {Becker}, \citenamefont {Schiemann}, \citenamefont {K{\"u}lz},\ and\ \citenamefont {Bergmann}}]{gaubatz_population_1988}%
  \BibitemOpen
  \bibfield  {author} {\bibinfo {author} {\bibfnamefont {U.}~\bibnamefont {Gaubatz}}, \bibinfo {author} {\bibfnamefont {P.}~\bibnamefont {Rudecki}}, \bibinfo {author} {\bibfnamefont {M.}~\bibnamefont {Becker}}, \bibinfo {author} {\bibfnamefont {S.}~\bibnamefont {Schiemann}}, \bibinfo {author} {\bibfnamefont {M.}~\bibnamefont {K{\"u}lz}},\ and\ \bibinfo {author} {\bibfnamefont {K.}~\bibnamefont {Bergmann}},\ }\bibfield  {title} {\bibinfo {title} {Population switching between vibrational levels in molecular beams},\ }\href {https://api.semanticscholar.org/CorpusID:96942434} {\bibfield  {journal} {\bibinfo  {journal} {Chemical Physics Letters}\ }\textbf {\bibinfo {volume} {149}},\ \bibinfo {pages} {463--468} (\bibinfo {year} {1988})}\BibitemShut {NoStop}%
\bibitem [{\citenamefont {Gaubatz}\ \emph {et~al.}(1990)\citenamefont {Gaubatz}, \citenamefont {Rudecki}, \citenamefont {Schiemann},\ and\ \citenamefont {Bergmann}}]{gaubatz1990population}%
  \BibitemOpen
  \bibfield  {author} {\bibinfo {author} {\bibfnamefont {U.}~\bibnamefont {Gaubatz}}, \bibinfo {author} {\bibfnamefont {P.}~\bibnamefont {Rudecki}}, \bibinfo {author} {\bibfnamefont {S.}~\bibnamefont {Schiemann}},\ and\ \bibinfo {author} {\bibfnamefont {K.}~\bibnamefont {Bergmann}},\ }\bibfield  {title} {\bibinfo {title} {Population transfer between molecular vibrational levels by stimulated raman scattering with partially overlapping laser fields. a new concept and experimental results},\ }\href {https://api.semanticscholar.org/CorpusID:97081066} {\bibfield  {journal} {\bibinfo  {journal} {Journal of Chemical Physics}\ }\textbf {\bibinfo {volume} {92}},\ \bibinfo {pages} {5363--5376} (\bibinfo {year} {1990})}\BibitemShut {NoStop}%
\bibitem [{\citenamefont {Shore}(2008)}]{Shore2008CoherentMO}%
  \BibitemOpen
  \bibfield  {author} {\bibinfo {author} {\bibfnamefont {B.~W.}\ \bibnamefont {Shore}},\ }\bibfield  {title} {\bibinfo {title} {Coherent manipulations of atoms using laser light},\ }\href {https://api.semanticscholar.org/CorpusID:120229675} {\bibfield  {journal} {\bibinfo  {journal} {Acta Physica Slovaca}\ }\textbf {\bibinfo {volume} {58}},\ \bibinfo {pages} {243--486} (\bibinfo {year} {2008})}\BibitemShut {NoStop}%
\bibitem [{\citenamefont {Shore}(2013)}]{shore2013}%
  \BibitemOpen
  \bibfield  {author} {\bibinfo {author} {\bibfnamefont {B.~W.}\ \bibnamefont {Shore}},\ }\bibfield  {title} {\bibinfo {title} {Pre-history of the concepts underlying stimulated raman adiabatic passage (stirap)},\ }\href {https://doi.org/10.2478/apsrt-2013-0006} {\bibfield  {journal} {\bibinfo  {journal} {Acta Physica Slovaca}\ }\textbf {\bibinfo {volume} {63}},\ \bibinfo {pages} {361--482} (\bibinfo {year} {2013})}\BibitemShut {NoStop}%
\bibitem [{\citenamefont {Bergmann}\ \emph {et~al.}(2015)\citenamefont {Bergmann}, \citenamefont {Vitanov},\ and\ \citenamefont {Shore}}]{bergmann2015perspective}%
  \BibitemOpen
  \bibfield  {author} {\bibinfo {author} {\bibfnamefont {K.}~\bibnamefont {Bergmann}}, \bibinfo {author} {\bibfnamefont {N.~V.}\ \bibnamefont {Vitanov}},\ and\ \bibinfo {author} {\bibfnamefont {B.~W.}\ \bibnamefont {Shore}},\ }\bibfield  {title} {\bibinfo {title} {Perspective: Stimulated raman adiabatic passage: The status after 25 years.},\ }\href {https://api.semanticscholar.org/CorpusID:205200242} {\bibfield  {journal} {\bibinfo  {journal} {The Journal of chemical physics}\ }\textbf {\bibinfo {volume} {142 17}},\ \bibinfo {pages} {170901} (\bibinfo {year} {2015})}\BibitemShut {NoStop}%
\bibitem [{\citenamefont {Vitanov}\ \emph {et~al.}(2017)\citenamefont {Vitanov}, \citenamefont {Rangelov}, \citenamefont {Shore},\ and\ \citenamefont {Bergmann}}]{vitanov2017}%
  \BibitemOpen
  \bibfield  {author} {\bibinfo {author} {\bibfnamefont {N.~V.}\ \bibnamefont {Vitanov}}, \bibinfo {author} {\bibfnamefont {A.~A.}\ \bibnamefont {Rangelov}}, \bibinfo {author} {\bibfnamefont {B.~W.}\ \bibnamefont {Shore}},\ and\ \bibinfo {author} {\bibfnamefont {K.}~\bibnamefont {Bergmann}},\ }\bibfield  {title} {\bibinfo {title} {Stimulated raman adiabatic passage in physics, chemistry, and beyond},\ }\href {https://doi.org/10.1103/RevModPhys.89.015006} {\bibfield  {journal} {\bibinfo  {journal} {Rev. Mod. Phys.}\ }\textbf {\bibinfo {volume} {89}},\ \bibinfo {pages} {015006} (\bibinfo {year} {2017})}\BibitemShut {NoStop}%
\bibitem [{\citenamefont {Golter}\ \emph {et~al.}(2013)\citenamefont {Golter}, \citenamefont {Dinyari},\ and\ \citenamefont {Wang}}]{golter2013nuclear}%
  \BibitemOpen
  \bibfield  {author} {\bibinfo {author} {\bibfnamefont {D.~A.}\ \bibnamefont {Golter}}, \bibinfo {author} {\bibfnamefont {K.~N.}\ \bibnamefont {Dinyari}},\ and\ \bibinfo {author} {\bibfnamefont {H.}~\bibnamefont {Wang}},\ }\bibfield  {title} {\bibinfo {title} {Nuclear-spin-dependent coherent population trapping of single nitrogen-vacancy centers in diamond},\ }\href {https://doi.org/10.1103/PhysRevA.87.035801} {\bibfield  {journal} {\bibinfo  {journal} {Phys. Rev. A}\ }\textbf {\bibinfo {volume} {87}},\ \bibinfo {pages} {035801} (\bibinfo {year} {2013})}\BibitemShut {NoStop}%
\bibitem [{\citenamefont {Qiao}\ \emph {et~al.}(2014)\citenamefont {Qiao}, \citenamefont {Ren}, \citenamefont {Chen}, \citenamefont {Bellaiche}, \citenamefont {Zhang}, \citenamefont {MacDonald},\ and\ \citenamefont {Niu}}]{golter2014optically}%
  \BibitemOpen
  \bibfield  {author} {\bibinfo {author} {\bibfnamefont {Z.}~\bibnamefont {Qiao}}, \bibinfo {author} {\bibfnamefont {W.}~\bibnamefont {Ren}}, \bibinfo {author} {\bibfnamefont {H.}~\bibnamefont {Chen}}, \bibinfo {author} {\bibfnamefont {L.}~\bibnamefont {Bellaiche}}, \bibinfo {author} {\bibfnamefont {Z.}~\bibnamefont {Zhang}}, \bibinfo {author} {\bibfnamefont {A.~H.}\ \bibnamefont {MacDonald}},\ and\ \bibinfo {author} {\bibfnamefont {Q.}~\bibnamefont {Niu}},\ }\bibfield  {title} {\bibinfo {title} {Quantum anomalous hall effect in graphene proximity coupled to an antiferromagnetic insulator},\ }\href {https://doi.org/10.1103/PhysRevLett.112.116404} {\bibfield  {journal} {\bibinfo  {journal} {Phys. Rev. Lett.}\ }\textbf {\bibinfo {volume} {112}},\ \bibinfo {pages} {116404} (\bibinfo {year} {2014})}\BibitemShut {NoStop}%
\bibitem [{\citenamefont {Golter}\ \emph {et~al.}(2014)\citenamefont {Golter}, \citenamefont {Baldwin},\ and\ \citenamefont {Wang}}]{golter2014protecting}%
  \BibitemOpen
  \bibfield  {author} {\bibinfo {author} {\bibfnamefont {D.~A.}\ \bibnamefont {Golter}}, \bibinfo {author} {\bibfnamefont {T.~K.}\ \bibnamefont {Baldwin}},\ and\ \bibinfo {author} {\bibfnamefont {H.}~\bibnamefont {Wang}},\ }\bibfield  {title} {\bibinfo {title} {Protecting a solid-state spin from decoherence using dressed spin states},\ }\href {https://doi.org/10.1103/PhysRevLett.113.237601} {\bibfield  {journal} {\bibinfo  {journal} {Phys. Rev. Lett.}\ }\textbf {\bibinfo {volume} {113}},\ \bibinfo {pages} {237601} (\bibinfo {year} {2014})}\BibitemShut {NoStop}%
\bibitem [{\citenamefont {Di~Stefano}\ \emph {et~al.}(2016)\citenamefont {Di~Stefano}, \citenamefont {Paladino}, \citenamefont {Pope},\ and\ \citenamefont {Falci}}]{di2016coherent}%
  \BibitemOpen
  \bibfield  {author} {\bibinfo {author} {\bibfnamefont {P.~G.}\ \bibnamefont {Di~Stefano}}, \bibinfo {author} {\bibfnamefont {E.}~\bibnamefont {Paladino}}, \bibinfo {author} {\bibfnamefont {T.~J.}\ \bibnamefont {Pope}},\ and\ \bibinfo {author} {\bibfnamefont {G.}~\bibnamefont {Falci}},\ }\bibfield  {title} {\bibinfo {title} {Coherent manipulation of noise-protected superconducting artificial atoms in the lambda scheme},\ }\href {https://doi.org/10.1103/PhysRevA.93.051801} {\bibfield  {journal} {\bibinfo  {journal} {Phys. Rev. A}\ }\textbf {\bibinfo {volume} {93}},\ \bibinfo {pages} {051801} (\bibinfo {year} {2016})}\BibitemShut {NoStop}%
\bibitem [{\citenamefont {Kumar}\ \emph {et~al.}(2016)\citenamefont {Kumar}, \citenamefont {Vepsäläinen}, \citenamefont {Danilin},\ and\ \citenamefont {Paraoanu}}]{kumar2016stimulated}%
  \BibitemOpen
  \bibfield  {author} {\bibinfo {author} {\bibfnamefont {K.~S.}\ \bibnamefont {Kumar}}, \bibinfo {author} {\bibfnamefont {A.}~\bibnamefont {Vepsäläinen}}, \bibinfo {author} {\bibfnamefont {S.}~\bibnamefont {Danilin}},\ and\ \bibinfo {author} {\bibfnamefont {G.~S.}\ \bibnamefont {Paraoanu}},\ }\bibfield  {title} {\bibinfo {title} {Stimulated {Raman} adiabatic passage in a three-level superconducting circuit},\ }\href {https://doi.org/10.1038/ncomms10628} {\bibfield  {journal} {\bibinfo  {journal} {Nature Communications}\ }\textbf {\bibinfo {volume} {7}},\ \bibinfo {pages} {10628} (\bibinfo {year} {2016})}\BibitemShut {NoStop}%
\bibitem [{\citenamefont {Xu}\ \emph {et~al.}(2016)\citenamefont {Xu}, \citenamefont {Song}, \citenamefont {Liu}, \citenamefont {Xue}, \citenamefont {Su}, \citenamefont {Deng}, \citenamefont {Tian}, \citenamefont {Zheng}, \citenamefont {Han}, \citenamefont {Zhong}, \citenamefont {Wang}, \citenamefont {Liu},\ and\ \citenamefont {Zhao}}]{xu2016coherent}%
  \BibitemOpen
  \bibfield  {author} {\bibinfo {author} {\bibfnamefont {H.~K.}\ \bibnamefont {Xu}}, \bibinfo {author} {\bibfnamefont {C.}~\bibnamefont {Song}}, \bibinfo {author} {\bibfnamefont {W.~Y.}\ \bibnamefont {Liu}}, \bibinfo {author} {\bibfnamefont {G.~M.}\ \bibnamefont {Xue}}, \bibinfo {author} {\bibfnamefont {F.~F.}\ \bibnamefont {Su}}, \bibinfo {author} {\bibfnamefont {H.}~\bibnamefont {Deng}}, \bibinfo {author} {\bibfnamefont {Y.}~\bibnamefont {Tian}}, \bibinfo {author} {\bibfnamefont {D.~N.}\ \bibnamefont {Zheng}}, \bibinfo {author} {\bibfnamefont {S.}~\bibnamefont {Han}}, \bibinfo {author} {\bibfnamefont {Y.~P.}\ \bibnamefont {Zhong}}, \bibinfo {author} {\bibfnamefont {H.}~\bibnamefont {Wang}}, \bibinfo {author} {\bibfnamefont {Y.-x.}\ \bibnamefont {Liu}},\ and\ \bibinfo {author} {\bibfnamefont {S.~P.}\ \bibnamefont {Zhao}},\ }\bibfield  {title} {\bibinfo {title} {Coherent population transfer between uncoupled or weakly coupled states in ladder-type superconducting qutrits},\ }\href
  {https://doi.org/10.1038/ncomms11018} {\bibfield  {journal} {\bibinfo  {journal} {Nature Communications}\ }\textbf {\bibinfo {volume} {7}},\ \bibinfo {pages} {11018} (\bibinfo {year} {2016})}\BibitemShut {NoStop}%
\bibitem [{\citenamefont {Premaratne}\ \emph {et~al.}(2017)\citenamefont {Premaratne}, \citenamefont {Wellstood},\ and\ \citenamefont {Palmer}}]{premaratne2017microwave}%
  \BibitemOpen
  \bibfield  {author} {\bibinfo {author} {\bibfnamefont {S.~P.}\ \bibnamefont {Premaratne}}, \bibinfo {author} {\bibfnamefont {F.~C.}\ \bibnamefont {Wellstood}},\ and\ \bibinfo {author} {\bibfnamefont {B.~S.}\ \bibnamefont {Palmer}},\ }\bibfield  {title} {\bibinfo {title} {Microwave photon {Fock} state generation by stimulated {Raman} adiabatic passage},\ }\href {https://doi.org/10.1038/ncomms14148} {\bibfield  {journal} {\bibinfo  {journal} {Nature Communications}\ }\textbf {\bibinfo {volume} {8}},\ \bibinfo {pages} {14148} (\bibinfo {year} {2017})}\BibitemShut {NoStop}%
\bibitem [{\citenamefont {Kandel}\ \emph {et~al.}(2021)\citenamefont {Kandel}, \citenamefont {Qiao}, \citenamefont {Fallahi}, \citenamefont {Gardner}, \citenamefont {Manfra},\ and\ \citenamefont {Nichol}}]{kandel2021}%
  \BibitemOpen
  \bibfield  {author} {\bibinfo {author} {\bibfnamefont {Y.~P.}\ \bibnamefont {Kandel}}, \bibinfo {author} {\bibfnamefont {H.}~\bibnamefont {Qiao}}, \bibinfo {author} {\bibfnamefont {S.}~\bibnamefont {Fallahi}}, \bibinfo {author} {\bibfnamefont {G.~C.}\ \bibnamefont {Gardner}}, \bibinfo {author} {\bibfnamefont {M.~J.}\ \bibnamefont {Manfra}},\ and\ \bibinfo {author} {\bibfnamefont {J.~M.}\ \bibnamefont {Nichol}},\ }\bibfield  {title} {\bibinfo {title} {Adiabatic quantum state transfer in a semiconductor quantum-dot spin chain},\ }\href {https://doi.org/10.1038/s41467-021-22416-5} {\bibfield  {journal} {\bibinfo  {journal} {Nature Communications}\ }\textbf {\bibinfo {volume} {12}},\ \bibinfo {pages} {2156} (\bibinfo {year} {2021})}\BibitemShut {NoStop}%
\bibitem [{\citenamefont {Breuer}\ and\ \citenamefont {Petruccione}(2002)}]{Petruccione}%
  \BibitemOpen
  \bibfield  {author} {\bibinfo {author} {\bibfnamefont {H.~P.}\ \bibnamefont {Breuer}}\ and\ \bibinfo {author} {\bibfnamefont {F.}~\bibnamefont {Petruccione}},\ }\href {https://doi.org/10.1093/acprof:oso/9780199213900.001.0001} {\emph {\bibinfo {title} {The Theory Of Open Quantum Systems}}}\ (\bibinfo  {publisher} {Oxford University Press, Oxford},\ \bibinfo {year} {2002})\BibitemShut {NoStop}%
\bibitem [{\citenamefont {Rivas}\ and\ \citenamefont {Huelga}(2012)}]{Rivas2011OpenQS}%
  \BibitemOpen
  \bibfield  {author} {\bibinfo {author} {\bibfnamefont {A.}~\bibnamefont {Rivas}}\ and\ \bibinfo {author} {\bibfnamefont {S.~F.}\ \bibnamefont {Huelga}},\ }\href {https://doi.org/10.1007/978-3-642-23354-8} {\emph {\bibinfo {title} {Open Quantum Systems: An Introduction}}}\ (\bibinfo  {publisher} {Springer Berlin Heidelberg},\ \bibinfo {year} {2012})\BibitemShut {NoStop}%
\bibitem [{\citenamefont {Garrison}(1986)}]{Garrison1986}%
  \BibitemOpen
  \bibfield  {author} {\bibinfo {author} {\bibfnamefont {C.}~\bibnamefont {Garrison}},\ }\href@noop {} {\bibfield  {journal} {\bibinfo  {journal} {Preprint UCRL 94267 Lawrence Livermore Laboratory}\ } (\bibinfo {year} {1986})}\BibitemShut {NoStop}%
\bibitem [{\citenamefont {Rigolin}\ \emph {et~al.}(2008)\citenamefont {Rigolin}, \citenamefont {Ortiz},\ and\ \citenamefont {Ponce}}]{Rigolin2008}%
  \BibitemOpen
  \bibfield  {author} {\bibinfo {author} {\bibfnamefont {G.}~\bibnamefont {Rigolin}}, \bibinfo {author} {\bibfnamefont {G.}~\bibnamefont {Ortiz}},\ and\ \bibinfo {author} {\bibfnamefont {V.~H.}\ \bibnamefont {Ponce}},\ }\bibfield  {title} {\bibinfo {title} {Beyond the quantum adiabatic approximation: Adiabatic perturbation theory},\ }\href {https://doi.org/10.1103/PhysRevA.78.052508} {\bibfield  {journal} {\bibinfo  {journal} {Phys. Rev. A}\ }\textbf {\bibinfo {volume} {78}},\ \bibinfo {pages} {052508} (\bibinfo {year} {2008})}\BibitemShut {NoStop}%
\bibitem [{\citenamefont {De~Grandi}\ and\ \citenamefont {Polkovnikov}(2010)}]{DeGrandi2010}%
  \BibitemOpen
  \bibfield  {author} {\bibinfo {author} {\bibfnamefont {C.}~\bibnamefont {De~Grandi}}\ and\ \bibinfo {author} {\bibfnamefont {A.}~\bibnamefont {Polkovnikov}},\ }\bibfield  {title} {\bibinfo {title} {Adiabatic {Perturbation} {Theory}: {From} {Landau}–{Zener} {Problem} to {Quenching} {Through} a {Quantum} {Critical} {Point}},\ }in\ \href {https://doi.org/10.1007/978-3-642-11470-0_4} {\emph {\bibinfo {booktitle} {Quantum {Quenching}, {Annealing} and {Computation}}}},\ \bibinfo {editor} {edited by\ \bibinfo {editor} {\bibfnamefont {A.~K.}\ \bibnamefont {Chandra}}, \bibinfo {editor} {\bibfnamefont {A.}~\bibnamefont {Das}},\ and\ \bibinfo {editor} {\bibfnamefont {B.~K.}\ \bibnamefont {Chakrabarti}}}\ (\bibinfo  {publisher} {Springer Berlin Heidelberg},\ \bibinfo {address} {Berlin, Heidelberg},\ \bibinfo {year} {2010})\ pp.\ \bibinfo {pages} {75--114}\BibitemShut {NoStop}%
\bibitem [{\citenamefont {Milton}\ and\ \citenamefont {Stegun}(1970)}]{Mathfunction}%
  \BibitemOpen
  \bibfield  {author} {\bibinfo {author} {\bibfnamefont {A.}~\bibnamefont {Milton}}\ and\ \bibinfo {author} {\bibfnamefont {I.~A.}\ \bibnamefont {Stegun}},\ }\href@noop {} {\emph {\bibinfo {title} {Hankbook of Mathematical Functions:with Formulas, Graphs, and Mathematical Tables}}}\ (\bibinfo  {publisher} {Dover Publications},\ \bibinfo {address} {New York},\ \bibinfo {year} {1970})\BibitemShut {NoStop}%
\bibitem [{\citenamefont {N.~V.~Vitanov}\ and\ \citenamefont {K.}(2001)}]{2001_Book_Vitanov}%
  \BibitemOpen
  \bibfield  {author} {\bibinfo {author} {\bibfnamefont {B.~W.~S.}\ \bibnamefont {N.~V.~Vitanov}, \bibfnamefont {M.~Fleischhauer}}\ and\ \bibinfo {author} {\bibfnamefont {B.}~\bibnamefont {K.}},\ }\href@noop {} {\emph {\bibinfo {title} {Advances in Atomic, Molecular, and Optical Physics}}}\ (\bibinfo  {publisher} {Academic Press},\ \bibinfo {year} {2001})\BibitemShut {NoStop}%
\bibitem [{\citenamefont {H.~Goldstein}(2001)}]{CMechanics}%
  \BibitemOpen
  \bibfield  {author} {\bibinfo {author} {\bibfnamefont {J.~S.}\ \bibnamefont {H.~Goldstein}, \bibfnamefont {C.~Poole}},\ }\href@noop {} {\emph {\bibinfo {title} {Classical Mechanics}}}\ (\bibinfo  {publisher} {Addison Wesley},\ \bibinfo {address} {San Francisco},\ \bibinfo {year} {2001})\BibitemShut {NoStop}%
\bibitem [{\citenamefont {Borceux}\ and\ \citenamefont {Janelidze}(2001)}]{Borceux_Janelidze_2001}%
  \BibitemOpen
  \bibfield  {author} {\bibinfo {author} {\bibfnamefont {F.}~\bibnamefont {Borceux}}\ and\ \bibinfo {author} {\bibfnamefont {G.}~\bibnamefont {Janelidze}},\ }\href@noop {} {\emph {\bibinfo {title} {Galois Theories}}},\ Cambridge Studies in Advanced Mathematics\ (\bibinfo  {publisher} {Cambridge University Press},\ \bibinfo {year} {2001})\BibitemShut {NoStop}%
\bibitem [{\citenamefont {Hall}\ \emph {et~al.}(2014)\citenamefont {Hall}, \citenamefont {Cresser}, \citenamefont {Li},\ and\ \citenamefont {Andersson}}]{Andersson2014}%
  \BibitemOpen
  \bibfield  {author} {\bibinfo {author} {\bibfnamefont {M.~J.~W.}\ \bibnamefont {Hall}}, \bibinfo {author} {\bibfnamefont {J.~D.}\ \bibnamefont {Cresser}}, \bibinfo {author} {\bibfnamefont {L.}~\bibnamefont {Li}},\ and\ \bibinfo {author} {\bibfnamefont {E.}~\bibnamefont {Andersson}},\ }\bibfield  {title} {\bibinfo {title} {Canonical form of master equations and characterization of non-markovianity},\ }\href {https://doi.org/10.1103/PhysRevA.89.042120} {\bibfield  {journal} {\bibinfo  {journal} {Phys. Rev. A}\ }\textbf {\bibinfo {volume} {89}},\ \bibinfo {pages} {042120} (\bibinfo {year} {2014})}\BibitemShut {NoStop}%
\bibitem [{\citenamefont {Laine}\ \emph {et~al.}(2012)\citenamefont {Laine}, \citenamefont {Luoma},\ and\ \citenamefont {Piilo}}]{Laine_2012}%
  \BibitemOpen
  \bibfield  {author} {\bibinfo {author} {\bibfnamefont {E.-M.}\ \bibnamefont {Laine}}, \bibinfo {author} {\bibfnamefont {K.}~\bibnamefont {Luoma}},\ and\ \bibinfo {author} {\bibfnamefont {J.}~\bibnamefont {Piilo}},\ }\bibfield  {title} {\bibinfo {title} {Local-in-time master equations with memory effects: applicability and interpretation},\ }\href {https://doi.org/10.1088/0953-4075/45/15/154004} {\bibfield  {journal} {\bibinfo  {journal} {Journal of Physics B: Atomic, Molecular and Optical Physics}\ }\textbf {\bibinfo {volume} {45}},\ \bibinfo {pages} {154004} (\bibinfo {year} {2012})}\BibitemShut {NoStop}%
\bibitem [{\citenamefont {Hush}\ \emph {et~al.}(2015)\citenamefont {Hush}, \citenamefont {Lesanovsky},\ and\ \citenamefont {Garrahan}}]{Garrahan2015}%
  \BibitemOpen
  \bibfield  {author} {\bibinfo {author} {\bibfnamefont {M.~R.}\ \bibnamefont {Hush}}, \bibinfo {author} {\bibfnamefont {I.}~\bibnamefont {Lesanovsky}},\ and\ \bibinfo {author} {\bibfnamefont {J.~P.}\ \bibnamefont {Garrahan}},\ }\bibfield  {title} {\bibinfo {title} {Generic map from non-lindblad to lindblad master equations},\ }\href {https://doi.org/10.1103/PhysRevA.91.032113} {\bibfield  {journal} {\bibinfo  {journal} {Phys. Rev. A}\ }\textbf {\bibinfo {volume} {91}},\ \bibinfo {pages} {032113} (\bibinfo {year} {2015})}\BibitemShut {NoStop}%
\bibitem [{\citenamefont {Caldeira}\ and\ \citenamefont {Leggett}(1985)}]{PhysRevA.31.1059}%
  \BibitemOpen
  \bibfield  {author} {\bibinfo {author} {\bibfnamefont {A.~O.}\ \bibnamefont {Caldeira}}\ and\ \bibinfo {author} {\bibfnamefont {A.~J.}\ \bibnamefont {Leggett}},\ }\bibfield  {title} {\bibinfo {title} {Influence of damping on quantum interference: An exactly soluble model},\ }\href {https://doi.org/10.1103/PhysRevA.31.1059} {\bibfield  {journal} {\bibinfo  {journal} {Phys. Rev. A}\ }\textbf {\bibinfo {volume} {31}},\ \bibinfo {pages} {1059--1066} (\bibinfo {year} {1985})}\BibitemShut {NoStop}%
\bibitem [{\citenamefont {Vitanov}\ \emph {et~al.}(1999)\citenamefont {Vitanov}, \citenamefont {Suominen},\ and\ \citenamefont {Shore}}]{vitanov1999creation}%
  \BibitemOpen
  \bibfield  {author} {\bibinfo {author} {\bibfnamefont {N.}~\bibnamefont {Vitanov}}, \bibinfo {author} {\bibfnamefont {K.}~\bibnamefont {Suominen}},\ and\ \bibinfo {author} {\bibfnamefont {B.}~\bibnamefont {Shore}},\ }\bibfield  {title} {\bibinfo {title} {Creation of coherent atomic superpositions by fractional stimulated raman adiabatic passage},\ }\href@noop {} {\bibfield  {journal} {\bibinfo  {journal} {Journal of Physics B: Atomic, Molecular and Optical Physics}\ }\textbf {\bibinfo {volume} {32}},\ \bibinfo {pages} {4535} (\bibinfo {year} {1999})}\BibitemShut {NoStop}%
\end{thebibliography}

\end{document}